\documentclass[trackchanges,twocolumn]{aastex701}
\usepackage{longtable}
\usepackage{graphicx}%
\usepackage{multirow}%
\usepackage{amsmath,amssymb,amsfonts}%
\usepackage{mathrsfs}%
\usepackage{xcolor}%
\usepackage{textcomp}%
\usepackage{booktabs}%
\usepackage{algorithm}%
\usepackage{algorithmicx}%
\usepackage{algpseudocode}%
\usepackage{listings}%
\usepackage{overpic}
\usepackage{threeparttable}
\usepackage{makecell}

\begin{document}

\title{EP250302a: violent shell collision in a soft-X-ray-selected GRB-like transient}

\correspondingauthor{Wei-Hua Lei, Dong Xu, Tao An, Jin-Jun Geng, Xiang-Yu Wang}
\email{leiwh@hust.edu.cn, dxu@nao.cas.cn, antao@shao.ac.cn, jjgeng@pmo.ac.cn, xywang@nju.edu.cn}

\author[orcid=0009-0002-7730-3985]{Shao-Yu Fu}
\altaffiliation{These authors contributed equally to this work.}
\affiliation{Department of Astronomy, School of Physics, Huazhong University of Science and Technology, Wuhan, 430074, China}
\email{syfu@hust.edu.cn}

\author[orcid=0000-0002-0170-0741]{Cui-Yuan Dai}
\altaffiliation{These authors contributed equally to this work.}
\affiliation{School of Astronomy and Space Science, Nanjing University, Nanjing, 210093, China}
\affiliation{Key Laboratory of Modern Astronomy and Astrophysics (Nanjing University), Ministry of Education, Nanjing, 210093, China}
\email{}

\author[orcid=0000-0002-7351-5801]{Ai-Ling Wang}
\altaffiliation{These authors contributed equally to this work.}
\affiliation{Key Laboratory of Particle Astrophysics, Institute of High Energy Physics, Chinese Academy of Sciences, Beijing, 100049, China}
\affiliation{Spallation Neutron Source Science Center, Dongguan, 523803, China}
\affiliation{Yunnan Observatories, Chinese Academy of Sciences, 396 Yangfangwang, Guandu District, Kunming 650216, China}
\email{}

\author[orcid=0000-0003-3440-1526]{Wei-Hua Lei}
\affiliation{Department of Astronomy, School of Physics, Huazhong University of Science and Technology, Wuhan, 430074, China}
\affiliation{Hubei Key Laboratory of Gravitation and Quantum Physics, Huazhong University of Science and Technology, Wuhan, Hubei, China}
\affiliation{Hubei Fundamental Research Center for Physics, Wuhan, Hubei, China}
\email{leiwh@hust.edu.cn}

\author[orcid=0000-0003-3257-9435]{Dong Xu}
\affiliation{Key Laboratory of Optical Astronomy, National Astronomical Observatories, Chinese Academy of Sciences, Beijing, 100012, China}
\email{dxu@nao.cas.cn}

\author[orcid=0000-0003-4341-0029]{Tao An}
\affiliation{Shanghai Astronomical Observatory, Chinese Academy of Sciences, Shanghai, 200030, China}
\affiliation{School of Astronomy and Space Science, University of Chinese Academy of Sciences, Beijing, 100049, China}
\email{antao@shao.ac.cn}

\author[orcid=0000-0001-9648-7295]{Jin-Jun Geng}
\affiliation{Purple Mountain Observatory, Chinese Academy of Sciences, Nanjing, 210023, China}
\email{jjgeng@pmo.ac.cn}

\author[orcid=0000-0002-5881-335X]{Xiang-Yu Wang}
\affiliation{School of Astronomy and Space Science, Nanjing University, Nanjing, 210093, China}
\affiliation{Key Laboratory of Modern Astronomy and Astrophysics (Nanjing University), Ministry of Education, Nanjing, 210093, China}
\email{xywang@nju.edu.cn}

\author[orcid=0009-0001-8155-7905]{Shuai-Qing Jiang}
\affiliation{Key Laboratory of Optical Astronomy, National Astronomical Observatories, Chinese Academy of Sciences, Beijing, 100012, China}
\affiliation{School of Astronomy and Space Science, University of Chinese Academy of Sciences, Beijing, 100049, China}
\email{}

\author[orcid=0000-0002-9022-1928]{Zi-Pei Zhu}
\affiliation{Key Laboratory of Optical Astronomy, National Astronomical Observatories, Chinese Academy of Sciences, Beijing, 100012, China}
\email{}

\author{Xing Liu}
\affiliation{Key Laboratory of Optical Astronomy, National Astronomical Observatories, Chinese Academy of Sciences, Beijing, 100012, China}
\affiliation{School of Astronomy and Space Science, University of Chinese Academy of Sciences, Beijing, 100049, China}
\email{}

\author{Jie An}
\affiliation{Key Laboratory of Optical Astronomy, National Astronomical Observatories, Chinese Academy of Sciences, Beijing, 100012, China}
\affiliation{School of Astronomy and Space Science, University of Chinese Academy of Sciences, Beijing, 100049, China}
\email{}

\author{Lin-Bo He}
\affiliation{Key Laboratory of Optical Astronomy, National Astronomical Observatories, Chinese Academy of Sciences, Beijing, 100012, China}
\affiliation{School of Astronomy and Space Science, University of Chinese Academy of Sciences, Beijing, 100049, China}
\email{}

\author[orcid=0000-0002-8402-3722]{Jun-Jie Jin}
\affiliation{Key Laboratory of Optical Astronomy, National Astronomical Observatories, Chinese Academy of Sciences, Beijing, 100012, China}
\email{}

\author{Yu Zhang}
\affiliation{Key Laboratory of Optical Astronomy, National Astronomical Observatories, Chinese Academy of Sciences, Beijing, 100012, China}
\email{}

\author{Jinlei Zhang}
\affiliation{Key Laboratory of Optical Astronomy, National Astronomical Observatories, Chinese Academy of Sciences, Beijing, 100012, China}
\email{}

\author{Zhou Fan}
\affiliation{Key Laboratory of Optical Astronomy, National Astronomical Observatories, Chinese Academy of Sciences, Beijing, 100012, China}
\email{}

\author{Xing Gao}
\affiliation{Xinjiang Astronomical Observatory, Chinese Academy of Sciences, Urumqi, 830011, China}
\email{}

\author{Abdusamatjan Iskandar}
\affiliation{Xinjiang Astronomical Observatory, Chinese Academy of Sciences, Urumqi, 830011, China}
\email{}

\author{Shahidin Yaqup}
\affiliation{Xinjiang Astronomical Observatory, Chinese Academy of Sciences, Urumqi, 830011, China}
\email{}

\author{Tu-Hong Zhong}
\affiliation{Xinjiang Astronomical Observatory, Chinese Academy of Sciences, Urumqi, 830011, China}
\email{}

\author{Ali Esamdin}
\affiliation{Xinjiang Astronomical Observatory, Chinese Academy of Sciences, Urumqi, 830011, China}
\email{}

\author{Chun-Hai Bai}
\affiliation{Xinjiang Astronomical Observatory, Chinese Academy of Sciences, Urumqi, 830011, China}
\email{}

\author{Yu Zhang}
\affiliation{Xinjiang Astronomical Observatory, Chinese Academy of Sciences, Urumqi, 830011, China}
\email{}

\author[orcid=0000-0002-3100-6558]{He Gao}
\affiliation{School of Physics and Astronomy, Beijing Normal University, Beijing, 100875, China}
\affiliation{Institute for Frontier in Astronomy and Astrophysics, Beijing Normal University, Beijing, 102206, China}
\email{}

\author[orcid=0000-0002-6299-1263]{Xue-Feng Wu}
\affiliation{Purple Mountain Observatory, Chinese Academy of Sciences, Nanjing, 210023, China}
\email{}

\author[orcid=0000-0002-7517-326X]{Daniele Bjørn Malesani}
\affiliation{Cosmic Dawn Center (DAWN), Denmark}
\affiliation{Niels Bohr Institute, University of Copenhagen, Jagtvej 128, 2200, Copenhagen, Denmark}
\affiliation{Department of Astrophysics/IMAPP, Radboud University, Nijmegen, 6525 AJ, The Netherlands}
\email{}

\author{Luca Izzo}
\affiliation{Osservatorio Astronomico di Capodimonte, INAF, Salita Moiariello 16, Napoli, 80131, Italy}
\email{}

\author[orcid=0000-0002-8775-2365]{R. A. J. Eyles-Ferris}
\affiliation{School of Physics and Astronomy, University of Leicester, University Road, LE1 7RH, Leicester, UK}
\email{}

\author[orcid=0000-0002-6950-4587]{A. Saccardi}
\affiliation{Université Paris-Saclay, Université Paris Cité, CEA, CNRS, AIM, 91191, Gif-sur-Yvette, France}
\affiliation{Centre national d’études spatiales (CNES), Paris, France}
\email{}

\author[orcid=0000-0002-0187-8873]{B. Schneider}
\affiliation{Aix Marseille University, CNRS, CNES, LAM, Marseille, France}
\email{}

\author[orcid=0000-0002-9408-1563]{J. Palmerio}
\affiliation{Université Paris-Saclay, Université Paris Cité, CEA, CNRS, AIM, 91191, Gif-sur-Yvette, France}
\email{}

\author[orcid=0000-0003-3274-6336]{N. R. Tanvir}
\affiliation{School of Physics and Astronomy, University of Leicester, University Road, LE1 7RH, Leicester, UK}
\email{}

\author{Alexei Pozanenko}
\affiliation{Space Research Institute of RAS, Profsoyuznaya, Profsoyuznaya st., 84/32, Moscow, 117997, Russia}
\affiliation{National Research University Higher School of Economics, Myasnitskaya st., 20, Moscow, 101000, Russia}
\email{}

\author{Nicolai Pankov}
\affiliation{Space Research Institute of RAS, Profsoyuznaya, Profsoyuznaya st., 84/32, Moscow, 117997, Russia}
\affiliation{National Research University Higher School of Economics, Myasnitskaya st., 20, Moscow, 101000, Russia}
\email{}

\author{A. S. Moskvitin}
\affiliation{Special Astrophysical Observatory, Russian Academy of Sciences, 369167 Nizhnii Arkhyz, Russia}
\email{}

\author{O. I. Spiridonova}
\affiliation{Special Astrophysical Observatory, Russian Academy of Sciences, 369167 Nizhnii Arkhyz, Russia}
\email{}

\author{O. A. Maslennikova}
\affiliation{Special Astrophysical Observatory, Russian Academy of Sciences, 369167 Nizhnii Arkhyz, Russia}
\email{}

\author{A. Volnova}
\affiliation{Space Research Institute of RAS, Profsoyuznaya, Profsoyuznaya st., 84/32, Moscow, 117997, Russia}
\email{}

\author{E. Klunko}
\affiliation{Institute of Solar-Terrestrial Physics, Russian Academy of Sciences (Siberian Branch), 664033 Irkutsk, Russia}
\email{}

\author{V. Rumyantsev}
\affiliation{Crimean Astrophysical Observatory, Russian Academy of Sciences, 298409 Nauchny, Russia}
\email{}

\author{A. Volvach}
\affiliation{Crimean Astrophysical Observatory, Russian Academy of Sciences, 298409 Nauchny, Russia}
\email{}

\author{L. Volvach}
\affiliation{Crimean Astrophysical Observatory, Russian Academy of Sciences, 298409 Nauchny, Russia}
\email{}

\author[orcid=0000-0002-3415-4636]{Toktarkhan Komesh}
\affiliation{Energetic Cosmos Laboratory, Nazarbayev University, Astana 010000, Kazakhstan}
\affiliation{Xinjiang Astronomical Observatory, Chinese Academy of Sciences, Urumqi, 830011, China}
\affiliation{Institute of Experimental and Theoretical Physics, Al-Farabi Kazakh National University, Almaty 050040, Kazakhstan}
\email{}

\author[orcid=0000-0001-5481-7727]{Ernazar Abdikamalov}
\affiliation{Energetic Cosmos Laboratory, Nazarbayev University, Astana 010000, Kazakhstan}
\affiliation{Department of Physics, Nazarbayev University, 53 Kabanbay Batyr ave, 010000 Astana, Kazakhstan}
\email{}

\author[orcid=0000-0001-7629-7099]{Dilda Berdikhan}
\affiliation{Energetic Cosmos Laboratory, Nazarbayev University, Astana 010000, Kazakhstan}
\email{}

\author{Zhanat Maksut}
\affiliation{Energetic Cosmos Laboratory, Nazarbayev University, Astana 010000, Kazakhstan}
\affiliation{Fesenkov Astrophysical Institute, 050020, Almaty, Kazakhstan}
\email{}

\author[orcid=0000-0002-5400-3261]{Yuan-Chuan Zou}
\affiliation{Department of Astronomy, School of Physics, Huazhong University of Science and Technology, Wuhan, 430074, China}
\email{}

\author[orcid=0009-0009-5012-9666]{Hong-Zhou Wu}
\affiliation{Department of Astronomy, School of Physics, Huazhong University of Science and Technology, Wuhan, 430074, China}
\email{}

\author{Yun-Wei Yu}
\affiliation{Institute of Astrophysics, Central China Normal University, 430079, Wuhan, China}
\email{}

\author{Rong-Feng Shen}
\affiliation{School of Physics and Astronomy, Sun Yat-Sen University, 519082, Zhuhai, China}
\email{}

\author[orcid=0000-0002-8614-8721]{Yi-Han Wang}
\affiliation{Department of Astronomy, University of Wisconsin, WI 53706, Madison, USA}
\email{}

\author[orcid=0000-0002-9615-1481]{Hui Sun}
\affiliation{Key Laboratory of Optical Astronomy, National Astronomical Observatories, Chinese Academy of Sciences, Beijing, 100012, China}
\email{}

\author[orcid=0000-0003-4111-5958]{Bin-Bin Zhang}
\affiliation{School of Astronomy and Space Science, Nanjing University, Nanjing, 210093, China}
\affiliation{Key Laboratory of Modern Astronomy and Astrophysics (Nanjing University), Ministry of Education, Nanjing, 210093, China}
\email{}

\author[orcid=0000-0002-8708-0597]{Liang-Duan Liu}
\affiliation{Institute of Astrophysics, Central China Normal University, 430079, Wuhan, China}
\affiliation{Education Research and Application Center, National Astronomical Data Center, 430079, Wuhan, China}
\affiliation{Key Laboratory of Quark and Lepton Physics (Central China Normal University), Ministry of Education, 430079, Wuhan, China}
\email{}

\author[orcid=0000-0001-5931-2381]{Ye Li}
\affiliation{Purple Mountain Observatory, Chinese Academy of Sciences, Nanjing, 210023, China}
\affiliation{School of Astronomy and Space Sciences, University of Science and Technology of China, Hefei, China}
\email{}

\author[orcid=0000-0002-7320-5862]{Valerio D'Elia}
\affiliation{Space Science Data Center (SSDC) - Agenzia Spaziale Italiana (ASI), Roma, 00133, Italy}
\email{}

\author[orcid=0000-0002-9393-8078]{Ruben Salvaterra}
\affiliation{INAF-Istituto di Astrofisica Spaziale e Fisica Cosmica di Milano, Via A. Corti 12, Milano 20133, Italy}
\email{}

\author[orcid=0000-0002-4036-7419]{Massimiliano De Pasquale}
\affiliation{University of Messina, MIFT Department, via F. S. D’Alcontres 31, Messina, 98166, Italy}
\email{}

\author[orcid=0000-0002-9725-2524]{Bing Zhang}
\affiliation{The Hong Kong Institute for Astronomy and Astrophysics, The University of Hong Kong, Hong Kong, China}
\affiliation{Department of Physics, The University of Hong Kong, Pokfulam Road, Hong Kong, China}
\email{}

\author[orcid=0000-0001-8266-3024]{Wei-Min Yuan}
\affiliation{Key Laboratory of Optical Astronomy, National Astronomical Observatories, Chinese Academy of Sciences, Beijing, 100012, China}
\affiliation{School of Astronomy and Space Science, University of Chinese Academy of Sciences, Beijing, 100049, China}
\email{}

\begin{abstract}

The Einstein Probe opens a previously unexplored soft X-ray window onto gamma-ray bursts, filling a critical observational gap in the soft X-ray coverage of their prompt emission. In this letter, we present EP250302a, a soft-X-ray-selected, GRB-like transient at $z=1.131$  detected by the Einstein Probe. Follow-up observations from X-ray to radio reveal a narrow X-ray flare at $\sim 1.1$\, ks and subsequent achromatic optical and X-ray rebrightening. These features challenge a standard single-component afterglow model and indicate the need for multiple ejecta components. A violent collision between a late relativistic shell and the decelerated leading blast wave provides a plausible interpretation: the flare arises from internal dissipation of the late ejecta, while the rebrightening is powered by the shocked emission produced in the collision. Quantitative modeling constrains the kinetic energy ratio between the late shell and the initial ejecta to $E_{\rm k,iso,2}/E_{\rm k,iso,1} \sim 5$ (with $E_{\rm k,iso,2} \sim 10^{53}$~erg and $E_{\rm k,iso,1} \sim 2\times10^{52}$~erg), as well as the Lorentz factor contrast to $\Gamma_{2,0}/\Gamma_{1,0} \approx 0.98$--$2.27$, required to reproduce the observed flare luminosity and rebrightening amplitude. Such an energetic late shell can be launched in a radiatively inefficient second episode of central-engine activity. Thanks to the well-sampled, early-time multiband coverage facilitated by the EP trigger, EP250302a provides a valuable case to test the physical connection between central-engine activity and shell collisions.

\end{abstract}


\keywords{\uat{High energy astrophysics}{739} --- \uat{X-ray transient sources}{1852}}


\section{Introduction}\label{intro}


Extragalactic Fast X-ray Transients (EFXTs) have emerged as a class of high-energy phenomena with immense research potential. These transients are typically characterized by a dramatic surge in X-ray emission lasting from several seconds to hundreds of seconds, followed by a rapid decay. However, due to past observational constraints, most known samples were identified through archival data mining of missions such as Chandra and XMM-Newton \citep[e.g.,][]{2022A&A...663A.168Q, 2023A&A...675A..44Q}. These sources generally lack timely multi-wavelength follow-up observations during their initial outburst, which greatly hinders our understanding of their burst mechanisms and physical origins. 

The Einstein Probe (EP) satellite, which commenced operations in early 2024, fills a critical observational gap with its wide-field soft X-ray monitoring capability (0.5–4 keV). Integrated with a real-time alert system based on the Beidou short-message service, EP offers an unprecedented opportunity to systematically capture the early-time emission features of EFXTs \citep{2022hxga.book...86Y}. During its two years of on-orbit operation, EP has detected numerous prototypical EFXTs: EP240219a was found to be associated with a sub-threshold gamma-ray burst (GRB) detected by Fermi-GBM, and EP240315a was associated with GRB 240315A, proving that a fraction of FXTs are soft X-ray counterparts of GRBs and revealing complex soft X-ray components in early GRB emission \citep{2024ApJ...975L..27Y, 2025NatAs...9..564L}. EP240414a revealed a new population of Wolf-Rayet star explosions characterized by ``weak-jet engines'', where the central engine power is low yet sufficient to drive a successful jet, thereby filling the gap between standard long GRBs and low-luminosity GRBs \citep{2025NatAs...9.1073S}. The spectral and light-curve characteristics of EP240408a do not fully match any known transient classes, potentially representing a rare jetted tidal disruption event (TDE) or an entirely new class of phenomena \citep{2025SCPMA..6819511Z, 2025ApJ...979L..30O}. EP240222a represents a rare instance where a tidal disruption event (TDE) involving an intermediate-mass black hole (IMBH) has been captured in the X-ray band \citep{2025arXiv250109580J,2025ApJ...985L..48Y}, while EP250702a represents the direct observation of an IMBH disrupting a compact object \citep{2026SciBu..71..538L}. EP241113a provides a definitive observational confirmation of the ``dirty fireball'' model, which was theoretically predicted several decades ago \citep{2026arXiv260326213D}. Preliminary statistical analyses indicate that approximately one-third of EP-detected sources are explicitly associated with known GRBs, while another subset exhibits properties akin to GRBs despite the absence of detected gamma-ray counterparts (Wu et al. 2026 in prep). 

A primary scientific objective of the Einstein Probe is to establish a soft-X-ray-selected sample of GRB-like transients. By preferentially identifying sources characterized by low bulk Lorentz factors and intrinsically soft spectra, soft-X-ray triggering effectively fills the observational gap in the softer energy bands—a window that remains largely inaccessible to traditional $\gamma$-ray monitors due to their low-energy thresholds. This capability is exemplified by EP240315a, which demonstrated that soft-X-ray triggers can effectively capture the complex prompt emission of high-redshift GRBs \citep{2025NatAs...9..564L}. Furthermore, the discovery of the fast X-ray transient EP240414a and its associated supernova, SN 2024gsa, established a direct link between such transients and weakly relativistic jets emerging from stripped-envelope core-collapse explosions \citep{2025NatAs...9.1073S}. These cases demonstrate that the EP soft-X-ray band affords a unique, early-time probe of the diverse central-engine and ejecta properties inherent to relativistic transients.

During the Swift era, approximately one-third of GRBs were found to exhibit early-time X-ray flares at $10^2$--$10^4$~s post-burst, with rapid variability ($\delta t/t \ll 1$) that is difficult to reconcile with an external shock origin \citep{2005Sci...309.1833B, 2005ApJ...631..429I, 2006ApJ...642..354Z}. These flares are typically interpreted as signatures of central-engine reactivation or persistent activity, driven by mechanisms such as fallback accretion onto a newly formed black hole or magnetic dipole radiation from a newborn magnetar \citep{2005Sci...309.1833B, 2006ApJ...642..354Z}. The chromatic behavior of these flares---prominent in the X-ray band but often lacking a simultaneous optical counterpart---has been extensively analyzed, serving as a primary diagnostic of internal dissipation associated with ongoing engine activity \citep{Falcone2006, 2007A&A...469L..13M, 2010A&A...519A...4U, 2015ApJ...803...10T}. Continuous energy injection from the central engine can significantly alter the dynamical evolution of the afterglow, causing the light curve to deviate from standard power-law decays and manifest as shallow plateaus or distinct rebrightenings \citep{2001ApJ...552L..35Z,2006MNRAS.369..197F}. \citet{Zhang2002} provided a systematic theoretical framework for various forms of energy injection and their corresponding light curve signatures. They noted that in scenarios involving kinetic-energy-dominated shell injection, if the relative velocity between the catching-up shells exceeds a critical threshold, it triggers a violent shell collision. This process leads to intense, short-term rebrightening in the light curve. 
Recently, \citet{2025ApJ...994....5A} invoked this mechanism to model the radio flare and multiwavelength afterglow of GRB 231117A, underscoring both the theoretical expectation and observational prevalence of such late-time injection events.

Nevertheless, a critical observational gap remains. In most previous cases, GRBs exhibit either an X-ray flare or an optical rebrightening, but rarely both \citep{Li2012, Liang2007}; alternatively, the available data lack the simultaneous, high-quality multi-wavelength coverage to investigate the details of the violent collision \citep{2025ApJ...994....5A}. Consequently, events in which a late-time X-ray flare and a distinct optical bump are jointly detected with sufficient temporal and spectral resolution remain essentially unexplored. A clear diagnostic requires a source for which the prompt phase has definitively ended, followed by a distinct, late X-ray flare and a delayed, achromatic multi-band rebrightening, all captured in a continuous, gapless multi-wavelength dataset.

In this Letter, we present Einstein Probe (EP) observations of EP250302a, a soft-X-ray-selected, GRB-like transient at $z=1.131$. Enabled by EP's rapid alert capability and subsequent extensive multi-wavelength follow-up, we identified a sharp, ``needle-like'' X-ray flare at $\sim 1.1$~ks post-burst, followed by a prominent, achromatic rebrightening in both the optical and X-ray bands. Through systematic multi-model testing and quantitative broadband modeling, we show that the exceptional data coverage of EP250302a provides stringent constraints on the physical mechanisms of a violent shell collision. With continuous multi-wavelength coverage from the trigger through the late-time afterglow, EP250302a provides an ideal case for testing the physical connection between a late internal X-ray flare and the subsequent delayed optical-to-X-ray rebrightening within a unified framework of late central-engine reactivation and shell collisions.

\begin{figure*}[htbp]
\gridline{\fig{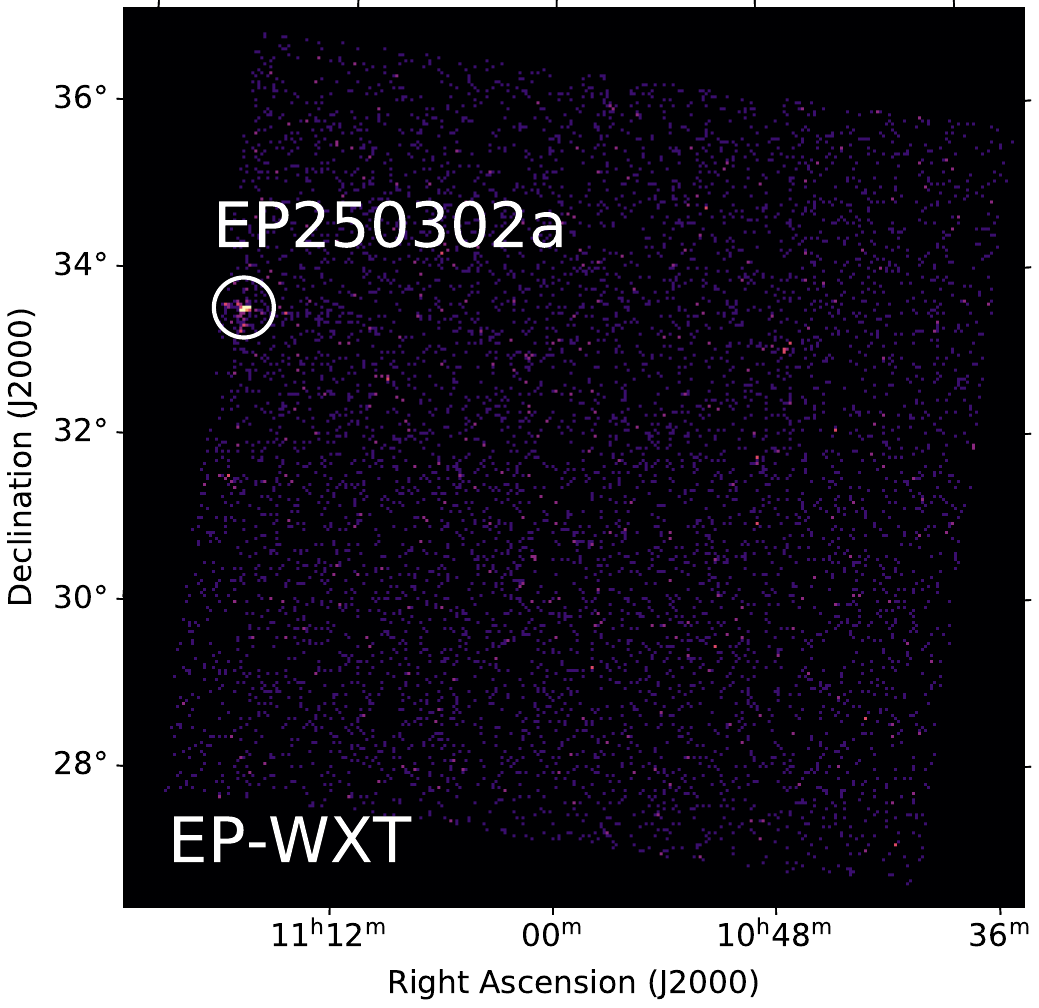}{0.42\textwidth}{(a)}
          \fig{skymap_FXT.pdf}{0.42\textwidth}{(b)}
          }
\caption{X-ray image of EP250302a. (a) EP-WXT image of EP250302a. (b) EP-FXT image of EP250302a obtained during the automatic follow-up observation. The outer yellow circle indicates the EP-WXT positional uncertainty with a radius of $2.24'$ (90\% confidence level). \label{fig_WXT_FXT_image}}
\end{figure*}

\section{Observation and Data Reduction}\label{obs}

\subsection{High Energy Observation}

At 15:36:04 (UTC) on 2025 March 2 ($T_0$), the Wide-field X-ray Telescope (WXT) onboard the Einstein Probe (EP) identified a novel X-ray transient, designated as EP250302a. The transient event lasted approximately 42\,s, reaching a peak flux of $9 \times 10^{-9} \text{ erg s}^{-1} \text{ cm}^{-2}$. Given the measured spectroscopic redshift of $z = 1.131$, the isotropic equivalent energy of the prompt emission at 0.5-4 keV is estimated to be $E_{\text{iso}} \sim 2 \times 10^{51} \text{ erg}$.

Following the trigger, the Follow-up X-ray Telescope (FXT) commenced autonomous observations within $\sim 100 \text{ s}$, revealing an uncatalogued source within the WXT error circle. The source is located at R.A. = $169.5151^{\circ}$, Dec. = $33.5851^{\circ}$ (J2000) with a positional uncertainty of 10". At the same time and position, \textit{Swift}/XRT also observed a fading X-ray source. EP-FXT continued monitoring the field in subsequent exposures (see Table~\ref{tab_FXT_obs}) and confirmed a monotonic decay of the source flux. Follow-up observations with \textit{Swift}/XRT and \textit{Chandra} continued for several days after the initial trigger.


We performed spectral analysis of the EP-WXT, EP-FXT, \textit{Swift}/XRT, and \textit{Chandra} data using \textit{XSPEC} v12.14.1. The spectra were fitted with an absorbed power-law model, \texttt{TBabs*zTBabs*cflux*powerlaw}, using C-statistics. We adopted \texttt{wilm} abundances and \texttt{vern} photoelectric cross sections. The Galactic absorption column density was fixed at $N_{\rm H,Gal}=2.3\times10^{20}\,{\rm cm^{-2}}$, obtained from the HEASARC \texttt{nH} tool based on the HI4PI H{\sc i} survey \citep{HI4PI2016}, and the redshift was fixed at $z=1.131$. The unabsorbed fluxes were calculated in the 0.5--4.0\,keV band for EP-WXT and in the 0.5--10.0\,keV band for EP-FXT, \textit{Swift}/XRT, and \textit{Chandra}.

For EP-FXT, source spectra were extracted from circular regions with a radius of $60^{\prime\prime}$ centered on the FXT position, while background spectra were extracted from nearby source-free circular regions with a radius of $180^{\prime\prime}$. The FXT data were divided into five phases based on the X-ray light curve for separate analysis: phase~1 (P1; 154--240\,s), phase~2 (P2; 240\,s--1.1\,ks), phase~3 (P3; 1.1--1.6\,ks), phase~4 (P4; 4.4--7.0\,ks), and phase~5 (P5; 44.7--47.8\,ks; Figure~\ref{fig_X-ray_lc}). The EP-FXT spectra contain approximately 36, 439, 1965, 1451, and 171 net counts in P1--P5, respectively. A later EP-FXT non-detection gives a 95\% confidence upper limit of $6.76\times10^{-15}\,{\rm erg\,cm^{-2}\,s^{-1}}$ in the 0.5--10.0\,keV band.

The \textit{Swift}/XRT event files were retrieved from the HEASARC archive and processed with XRTDAS. The \textit{Swift}/XRT spectra were extracted using \texttt{xrtproducts}, adopting a $60^{\prime\prime}$ source region and a nearby source-free background region. The two early \textit{Swift}/XRT spectra contain approximately 496 and 68 net counts, respectively, while the late-time combined \textit{Swift}/XRT spectrum contains approximately 14 net counts in the fitted energy range. A later \textit{Swift}/XRT non-detection at $\sim T_0+7$\,d gives a 90\% confidence upper limit of $1.18\times10^{-13}\,{\rm erg\,cm^{-2}\,s^{-1}}$ in the 0.3--10.0\,keV band.

The \textit{Chandra} data were reduced with CIAO v4.16.0 and CALDB v4.11.3. The \textit{Chandra} spectrum was extracted from a circular source region with a radius of $5^{\prime\prime}$ centered on the X-ray counterpart, with the background taken from a nearby source-free region. In the fitted energy range, the \textit{Chandra} spectrum contains 23 source-region counts, corresponding to $\simeq 22$ net counts after background subtraction, with a Li \& Ma detection significance of $\simeq 10\sigma$. For the low-count late-time spectra, we fixed $N_{\rm H,int}$ to $0.26\times10^{22}\,{\rm cm^{-2}}$, as inferred from the brighter joint EP-FXT/\textit{Swift}-XRT fit. The results of these analyses are summarized in Table~\ref{tab_spectrum_fitting}.

Utilizing the rapidGBM tool \citep{2025ApJ...993...51W}, we examined the Fermi spacecraft's orientation at the time of the EP250302a trigger and found that the source region was occulted by the Earth. Similarly, according to the Swift Mission Operations Center\footnote{\url{https://www.swift.psu.edu/operations/obsSchedule.php}}, the target location was outside the field of view of the Burst Alert Telescope (BAT). No significant emission was detected by INTEGRAL/SPI-ACS (80 keV--10 MeV) during the prompt phase, despite the source being within its field of view \citep{2025GCN.39907....1M}.


\subsection{Optical Observation}
\subsubsection{Photometry}
After receiving the trigger message, ground-based optical facilities automatically slewed to the source coordinates. The Nazarbayev University Transient Telescope at Assy-Turgen Astrophysical Observatory (NUTTelA-TAO) \citep[see][]{2023MNRAS.520.6104K} began optical observations of EP250302a 399\,s after the WXT trigger. Initial imaging was obtained in the Sloan $r'$ band until $t \approx 1.7$\,h after the trigger, after which simultaneous observations in the $g'$ and $r'$ bands were carried out, for a total observing duration of $\sim$2.4\,h. The half-meter telescope (HMT) at Nanshan Station and the AZT-33IK 1.5-meter telescope of Mondy Observatory automatically slewed to the burst position and obtained early optical light-curve coverage from $T_0+12.2$ min to $T_0+6.5$ hrs. A varying optical counterpart was identified within the FXT error circle, centered at R.A. = 11:18:03.58, Dec. = +33:35:09.06 (J2000) with a positional uncertainty of 1". The PAT17 optical telescope at Nanshan Station was manually triggered and also obtained the early light curve of the burst. The 2.6-meter telescope of the Crimean Astrophysical Observatory (CrAO) and the 1-meter Zeiss-1000 telescope of the Special Astrophysical Observatory of the RAS (SAO RAS), the 1.5-meter telescope of Sayan Solar Observatory (Mondy), and the Tsinghua-NAOC 0.8-m telescope (TNT) located at Xinglong, Hebei, China, conducted follow-up observations of this target. 

Swift/UVOT also monitored EP250302a in the $U$ band. We retrieved the UVOT sky images from the HEASARC archive and processed them with the UVOT tools in HEASoft, using the most recent CALDB. Individual exposures in each observing epoch were aspect-corrected and co-added with \texttt{uvotimsum}. Aperture photometry was then performed with \texttt{uvotsource}, using a circular source aperture with a radius of $5''$ centered on the optical counterpart and a nearby source-free background region. The count rates were converted to flux densities using the UVOT photometric calibration \citep{2008MNRAS.383..627P, 2011AIPC.1358..373B}. The UVOT $U$-band magnitudes reported in Table 6 are in the AB system. For epochs without a significant detection, we quote $3\sigma$ upper limits. No host-galaxy template subtraction was applied, given the absence of a detected host in archival images.

The ground-based optical data were reduced with IRAF, including bias and dark subtraction, flat-field correction, and combination to enhance the signal-to-noise ratio. Aperture photometry was performed on all ground-based images with Source Extractor \citep[version 2.28.2,][]{1996A&AS..117..393B} using an aperture diameter of $2\times$FWHM. The ground-based optical photometry was calibrated against Legacy Survey DR10 in the AB system. All the photometry results are listed in Table \ref{tab_phot}.

We also queried publicly available archival optical survey data and found no evidence of a host galaxy at the position of the optical counterpart. The deepest constraints are provided by the Legacy Survey, with non-detection limits of $g > 24.6$, $r > 24.0$, and $z > 23.6$ mag.

\subsubsection{Spectroscopy} 

We performed spectroscopic follow-up observations of EP250302a using the Beijing Faint Object Spectrograph and Camera (BFOSC) mounted on the Xinglong 2.16m telescope (Xinglong Observatory, Hebei, China). Observations started at 2025-03-02T16:27:24 UT, approximately 53 minutes after the burst trigger. We obtained three 1800\,s exposures with coverage of 3800 to 9000\,{\AA} using the G4 grism and 385LP filter; the flux standard star, Feige 34, was observed with the same configuration for 600\,s. Data reduction was conducted using standard IRAF procedures. However, the target remained too faint to yield any discernible emission or absorption features.

We also observed the fast X-ray transient EP250302a using the Multi Unit Spectroscopic Explorer (MUSE) integral-field spectrograph on the ESO VLT UT4 (Yepun). Observations commenced at a mid-epoch of $T+13.03$ hr, with a total integration time of 76.5 min. The raw data were reduced using the standard ESO MUSE pipeline \citep[version 2.10.14,][]{Weilbacher2020} executed within the ESO Reflex environment \citep{Freudling2013}. The standard reduction cascade was applied, which includes bias subtraction, flat-fielding, wavelength calibration, and illumination correction. The flux calibration and telluric absorption correction were performed using the observations of standard stars taken during the same night. After the basic calibration, the individual exposures were combined and resampled into a single 3D data cube. To further optimize the sky subtraction and remove complex sky emission line residuals, we applied the principal component analysis (PCA) based Zurich Atmosphere Purge \citep[ZAP,][]{Soto2016} algorithm to the final data cube. The fully calibrated data cube was then used to extract the local spectrum at the transient position and the nearby galaxy. The resulting continuous spectrum spans the wavelength range 4750–9300\,{\AA}. We identified a series of prominent absorption features at the blue end of the spectrum, whereas the redward portion is dominated by strong sky emission lines. The absorption features include $\text{Mg I}_{\lambda 2853}$, $\text{Mg II}_{\lambda 2796, 2804}$, and $\text{Fe II}_{\lambda 2344, 2374, 2383, 2587, 2600}$, all yielding a common redshift of $z = 1.131$, which is consistent with the results from VLT/X-shooter \citep{2025GCN.39574....1Y, 2025ApJ...993L..37O}. The features of the intervening system at $z = 0.549$ \citep{2025GCN.39561....1Y, 2025ApJ...993L..37O} fall in the blue part of the spectrum and are not covered by our data. As we do not detect any fine-structure lines, the measured redshift is formally a lower limit, as there could in principle be a weaker, unseen absorption system at high redshift. We nonetheless assume $z = 1.131$ in the following analysis. Leveraging the $1' \times 1'$ field of view of MUSE, we simultaneously extracted spectra for two neighboring galaxies. Both galaxies were determined to be at a redshift of $z = 0.127$, confirming that these sources are foreground objects and are not physically associated with EP250302a.

These two spectra from the Xinglong 2.16m telescope and VLT/MUSE are plotted in Figure \ref{EP250302a_spec}.

\begin{figure}[htbp]
\plotone{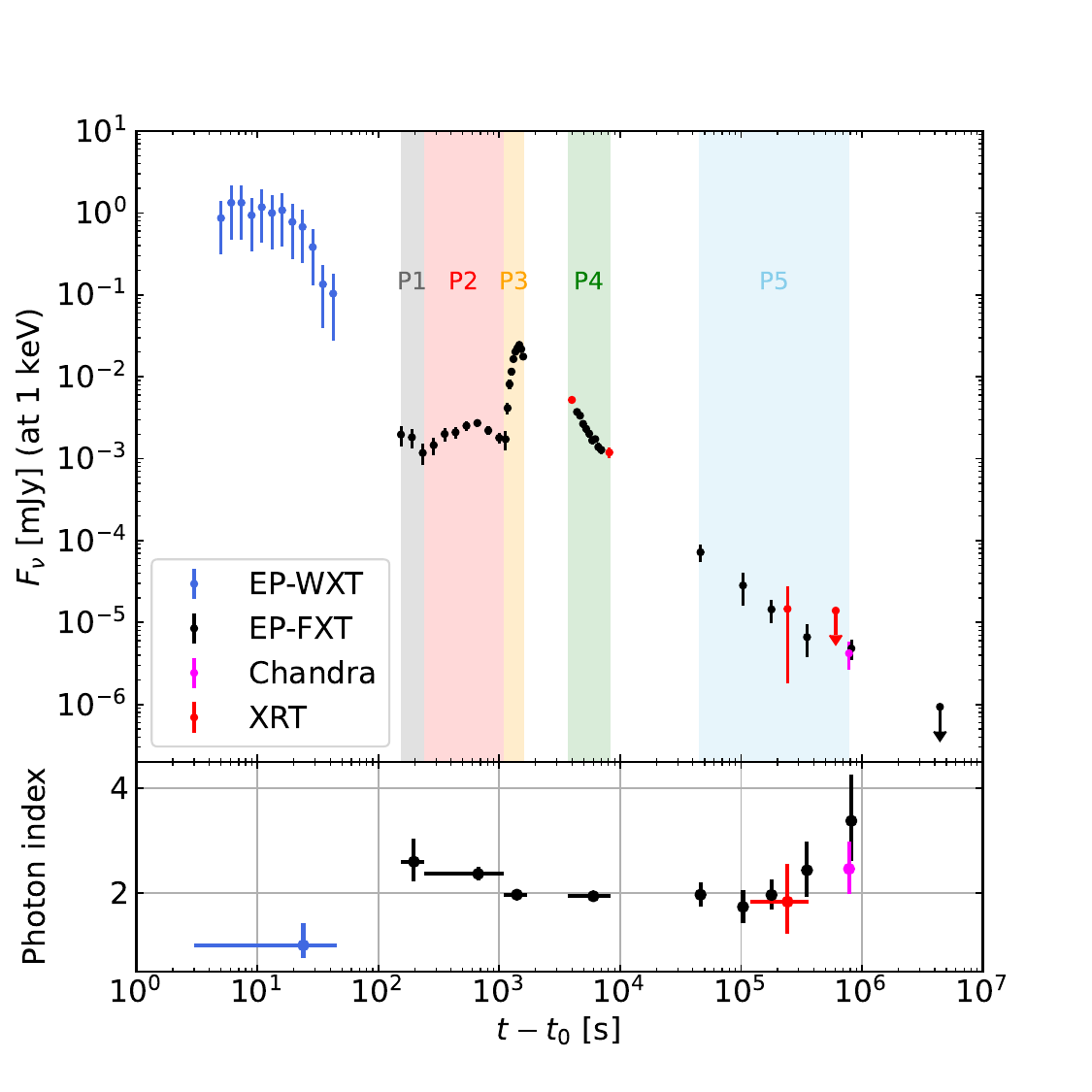}
\caption{The X-ray light curve of EP250302a. 
P1--P5 denote different time intervals (see text for details). The EP-WXT and EP-FXT light curves (during P1--P4) at 1\,keV are extrapolated from the count rate using the count-to-flux conversion factor derived from the best-fit average spectrum with an absorbed power-law model (Table~\ref{tab_spectrum_fitting}). For P5, the intrinsic absorption $N_{\rm int}$ is assumed to be constant across all data in this phase, while the photon index is allowed to vary. \label{fig_X-ray_lc}}
\end{figure}

\subsection{Radio Observation and Data Reduction}

\subsubsection{VLA} \label{sec:vla_obs}



We carried out three epochs of Karl G. Jansky Very Large Array continuum follow-up observations of EP250302a under program 25A-421 (PI: Ailing Wang) on 2025 March 15, 20, and 27. The observations were obtained in the D configuration.

The visibility data were calibrated in CASA v6.4.1 using the NRAO VLA calibration pipeline workflow. The pipeline applied standard a priori flagging and delay, bandpass, flux-density scale, and complex gain calibrations. The absolute flux-density scale and bandpass response were set using 3C286, and J1130+3815 was used as the complex gain calibrator. We inspected the pipeline weblogs and calibrated visibilities, and performed additional flagging of residual radio-frequency interference where necessary.

Continuum images were produced from the calibrated measurement sets using
CASA \texttt{tclean} with multi-frequency synthesis (\texttt{specmode=`mfs'}), Briggs weighting with \texttt{robust = 0.5}, \texttt{deconvolver=`multiscale'}, and \texttt{gridder=`standard'}. We imaged Stokes $I$ with an image size of $2048 \times 2048$ pixels. The deconvolution was run with \texttt{niter = 20000}, \texttt{gain = 0.1}, and a cleaning threshold of 0.001 mJy. The \texttt{nterms} parameter was not explicitly set, and the CASA default value was used. The final images were not primary-beam corrected during \texttt{tclean} (\texttt{pbcor=False}). The local rms noise was measured in source-free regions close to the target position.

Flux densities were measured at the known optical/X-ray afterglow position. Since the source is unresolved in all epochs, we adopt the peak flux density as the point-source flux density. As a consistency check, we fitted an unresolved Gaussian component using CASA \texttt{imfit}; the fitted integrated flux densities are consistent with the peak intensities within the local image rms. The quoted uncertainties are the local rms noise measured in source-free regions close to the target position. The resulting image-plane signal-to-noise ratios are approximately 3.4--4.5. The fitted radio positions are consistent with the optical/X-ray afterglow position within 3--5 arcsec.

\subsubsection{CrAO-RT22}
We conducted follow-up radio observations of the transient EP250302a using the 22m radio telescope (RT-22) located at the Crimean Astrophysical Observatory (CrAO) in Simeiz. The observations were performed at a center frequency of 36.8 GHz over three consecutive nights from 2025 March 3 to March 5. The source was not detected in any of the individual epochs. For each observation, the flux density was consistent with the background noise level. Consequently, we derived the $3\sigma$ upper limits for the radio emission at 36.8 GHz as follows: $3.3\times 10^{-4}$ Jy at $T_0+1.27$ days, $3.6\times 10^{-4}$ Jy at $T_0+2.28$ days, and $3.3\times 10^{-4}$ Jy at $T_0+3.29$ days, respectively.

\begin{figure*}[htbp]
	\centering
	\plotone{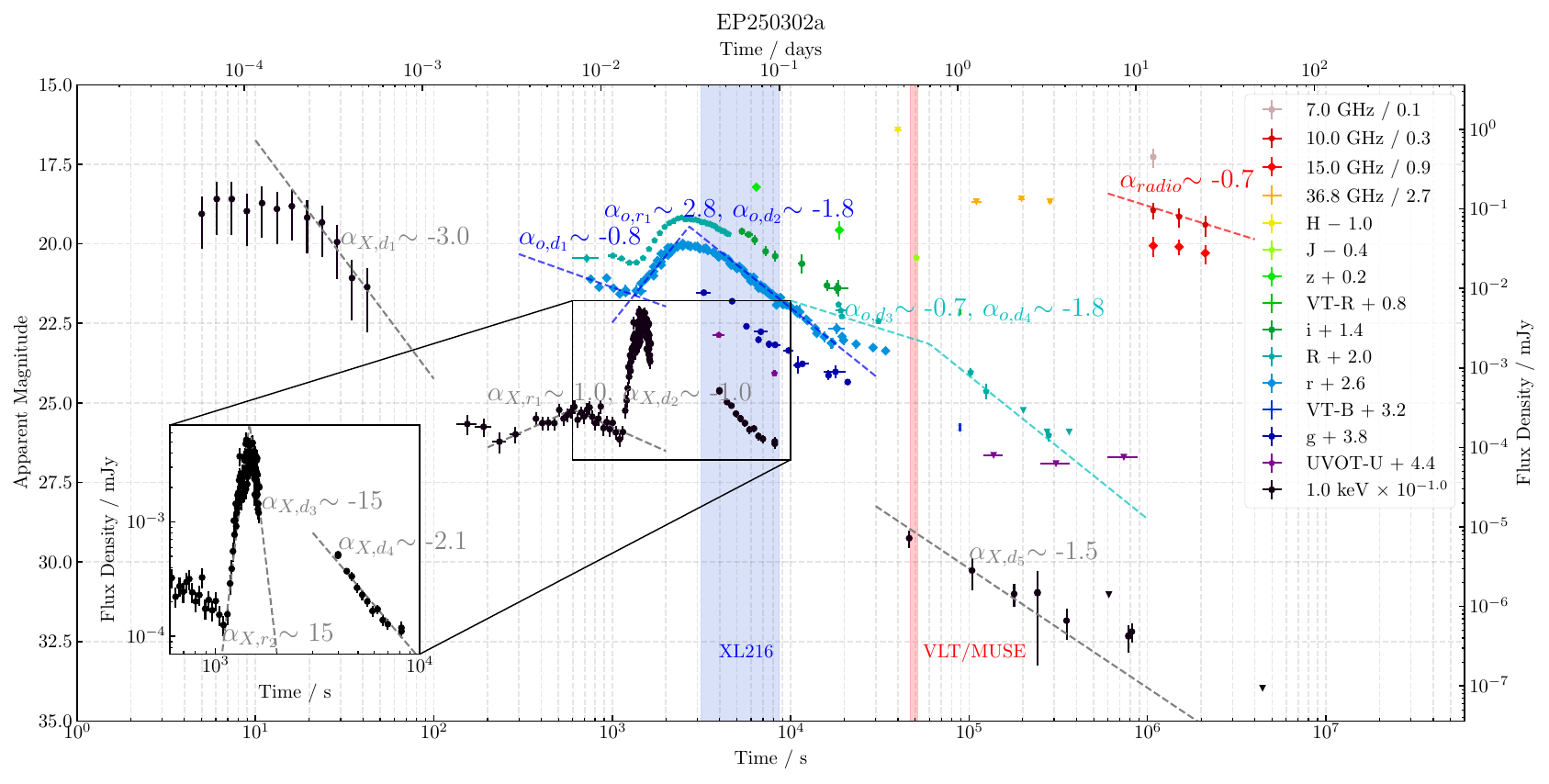}
	\caption{Multi-band light curve of EP250302a. The blue and red areas represent the spectroscopic observation time range of the Xinglong-2.16m telescope and VLT/MUSE. The rise/decay indexes of different band light curves are superimposed on the data points in the form of dashed lines.}
	\label{EP250302a_lc}
\end{figure*}

\section{Modeling and Results}

\subsection{Temporal Analysis}\label{mod}

The multi-band light curves at different phases are fit with a power law: 
\begin{equation}
    F(t) = A\left(\frac{t}{t_0}\right)^{\alpha},
\end{equation}
or a broken power law:
\begin{equation}
F(t) = 
\begin{cases}
    A(t/t_{\rm b})^{\alpha_1} & \text{if } t < t_{\rm b} \\
    A(t/t_{\rm b})^{\alpha_2} & \text{if } t \ge t_{\rm b} ,
\end{cases}
\end{equation}
where $t_0$ is the reference time of amplitude $A$ and $t_{\rm b}$ is the break time. The temporal index obtained from the fitting is labeled in Figure \ref{EP250302a_lc}.

For the multiband light curve during the giant optical bump, we adopt a smoothly broken power law to jointly fit the temporal and spectral behavior, assuming the SED between optical and X-ray can be described as a single power law 
\begin{equation}\label{eq.sbpl}
    F(t,\nu) = A_0\nu^{\beta_{\rm OX}} \left(\frac{t}{t_b}\right)^{\alpha_1}\left\{\frac{1}{2}\left[1+\left(\frac{t}{t_b}\right)^{1/\Delta}\right]\right\}^{(\alpha_2-\alpha_1)\Delta},
\end{equation}
where $\Delta$ is the smoothness parameter, and $\beta$ is the spectral index. We performed temporal and spectral fitting to the observational data using the MCMC method based on the equations described above. The temporal index for each segment is plotted in Figure \ref{EP250302a_lc}. Before performing spectral fitting, we corrected the optical data for Galactic extinction using the F19 model \citep{Fitzpatrick2019}. The spectral fitting results indicate that the spectrum is well described by a single power-law function, with a spectral index of $\beta_{\rm OX}=-0.70\pm0.01$.

Following the cessation of the prompt emission phase (see Figure \ref{EP250302a_lc}), the X-ray flux undergoes a steep decay with an index of $\alpha_{\rm{X},d_1} \approx -3.0$, smoothly transitioning into the afterglow phase. At the onset of the afterglow, the X-ray emission initially increases with an index of $\alpha_{\rm{X},r_1} \approx 1.0$, peaking at a few hundred seconds and subsequently decaying with $\alpha_{\rm{X},d_2} \approx -1.0$ until $\sim 1.1\,\rm{ks}$. This temporal behavior is broadly consistent with the initial optical decay index, $\alpha_{\rm{o},d_1} \approx -0.8$. Subsequently, a giant X-ray flare emerges at $\sim 1.1\,\rm{ks}$, peaking at $\sim 1.5\,\rm{ks}$ with rapid rise and decay indices of $\alpha_{\rm{X},r_2} \approx 15.0$ and $\alpha_{\rm{X},d_3} \approx -15.0$, respectively; notably, this feature is absent in the optical band. Instead, an optical rebrightening (bump) begins at $\sim 1.3\,\rm{ks}$—approximately $200\,\rm{s}$ after the onset of the X-ray flare—rising with $\alpha_{\rm{o},r_1} \approx 2.8$ and peaking at $\sim 2.5\,\rm{ks}$ before decaying with $\alpha_{\rm{o},d_2} \approx -1.8$. During this stage, the optical decay is consistent with the simultaneous X-ray decay index of $\alpha_{\rm{X},d_4} \approx -2.1$. Following this rebrightening phase, the optical light curve enters a shallow decay phase with $\alpha_{\rm{o},d_3} \approx -0.7$, before transitioning to a steeper decay ($\alpha_{\rm{o},d_4} \approx -1.8$) at $\sim 15\,\rm{ks}$. Meanwhile, the X-ray flux continues to decay with an index of $\alpha_{\rm{X},d_5} \approx -1.5$. At late times ($t \gtrsim 10$ days), the radio emission exhibits a persistent shallow decay consistent with an index of $\alpha_{\rm{radio}} \approx -0.7$.

\subsection{Theoretical Modelling}

The discrepancy between the onset times of the X-ray flare and the optical bump, coupled with their distinct light-curve morphologies, suggests that these emissions arise from two separate physical mechanisms. The rapid temporal evolution of the X-ray emission ($\Delta T/T \sim 0.5$), characterized by a fast rise and decay, is commonly seen in early GRB afterglows, often attributed to late-time activity of the central engine \citep{2006ApJ...642..389N, Falcone2006, 2007A&A...469L..13M, 2010A&A...519A...4U, 2015ApJ...803...10T}. Spectral modeling indicates that the flare emission is well-characterized by a single power-law (SPL) distribution with photon index $\Gamma_{\rm X}=1.95 \pm 0.04$. However, if this SPL component were to extend into the optical regime, a corresponding flux enhancement should be detectable in the optical band, which is inconsistent with our observations. The absence of such a signature in our data indicates that the X-ray flare likely originates from an internal region (a late-time injected shell) distinct from the site of the optical emission \citep{Falcone2006, 2007A&A...469L..13M, 2010A&A...519A...4U, 2015ApJ...803...10T}.

Approximately $200$\,s (in the observer frame) after the onset of the X-ray flare, the optical emission began to brighten. This rebrightening feature is observed across multiple optical bands, and is also accompanied by a contemporaneous bump component in the X-ray band. The multi-band temporal evolution of this component is well-described by a smoothly broken power law (Equation \ref{eq.sbpl}) with a single optical-to-X-ray spectral index $\beta_{\rm OX}= -0.70\pm 0.01$. If this feature were produced by continuous energy injection into the external forward shock (FS), the afterglow would be expected to maintain a consistent decay index $\alpha_{\rm o,d2}\approx -1.8$ at later times. However, the optical emission transitions to a shallower decay $\alpha_{\rm o,d3}\approx -0.7$ at $\sim 20\,$ks, suggesting that the physical origin of this multi-band giant rebrightening cannot be explained by a standard energy injection mechanism (or mild shell collision). Recent high-cadence optical observations of GRB 250702F have revealed a hybrid electron distribution at the forward shock, comprising a thermal (Maxwellian) component alongside a non-thermal power-law tail, as predicted by particle-in-cell simulations of ultra-relativistic collisionless shocks \citep{Jelinek2026}. This hybrid population also gives rise to an optical rebrightening in the afterglow light curve, phenomenologically consistent with the optical bump observed in EP250302a. Nevertheless, while the rebrightening in GRB 250702F is attributed to forward-shock electron thermalization, the prompt emission, X-ray flare, and optical bump in EP250302a are naturally unified within the violent shell collision scenario, providing a more comprehensive and physically motivated explanation for the multi-wavelength evolutionary behavior.

\subsubsection{Violent Collision of Two Shells}\label{viocoll}

The interval P2 (see Figure \ref{fig_X-ray_lc}) of the X-ray light curve is the afterglow emission of the outflow launched during the prompt phase. In the standard afterglow framework, the peak time of the P2 period corresponds to the dynamic deceleration time of the leading shock with an initial bulk Lorentz factor of $\Gamma_{1,0}$, marked as $T_{\rm dec,1}$. Beyond $T_{\rm dec,1}$ (or $R > R_{\rm dec,1} \simeq 2 \Gamma_{1,0}^2 c T_{\rm dec,1}$), the shock radius evolves as $R \propto t_{\rm obs}^{1/(4-k)}$ for a decelerating shell with self-similar evolution \citep{Blandford76} and $k$ is the profile index of the medium density of the circum-burst environment ($\propto R^{-k}$).
Assuming that another shell is launched from the central engine at the time of $T_0 + \Delta T$ after the burst trigger, it would catch up with the leading shock at the collision radius $R_{\rm coll}$~\citep{Geng25a}. The collision conditions for the two shells are
\begin{eqnarray}
    R_{\rm coll} &=& 2 \Gamma_{1,0}^2 c T_{\rm dec,1} \left(\frac{t_{\rm coll}}{T_{\rm dec,1}} \right)^{1/(4-k)}, \\
    R_{\rm coll} &=& 2 \Gamma_{2,0}^2 c (t_{\rm coll} - \Delta T)
\end{eqnarray}
respectively; combining them yields
\begin{equation}\label{eq.tcoll}
    \frac{\Gamma_{2,0}}{\Gamma_{1,0}} = \left(\frac{T_{\rm dec,1}}{t_{\rm coll} - \Delta T} \right)^{1/2} \left( \frac{t_{\rm coll}}{T_{\rm dec,1}} \right)^{1/(8-2k)}.
\end{equation}
For $T_{\rm dec,1} \simeq 400$~s, $\Delta T \simeq 1000$~s (in the observer frame) as inferred from the X-ray light curve, and assuming an ISM environment ($k = 0$), it gives $\Gamma_{2,0}/\Gamma_{1,0} \in [0.98, 2.27]$ for such a collision occurring within the interval P3, i.e., $t_{\rm coll} \in [1.1, 1.6]$~ks (see Figure \ref{fig_X-ray_lc}). 

\begin{figure}[htbp]
	\centering
	\plotone{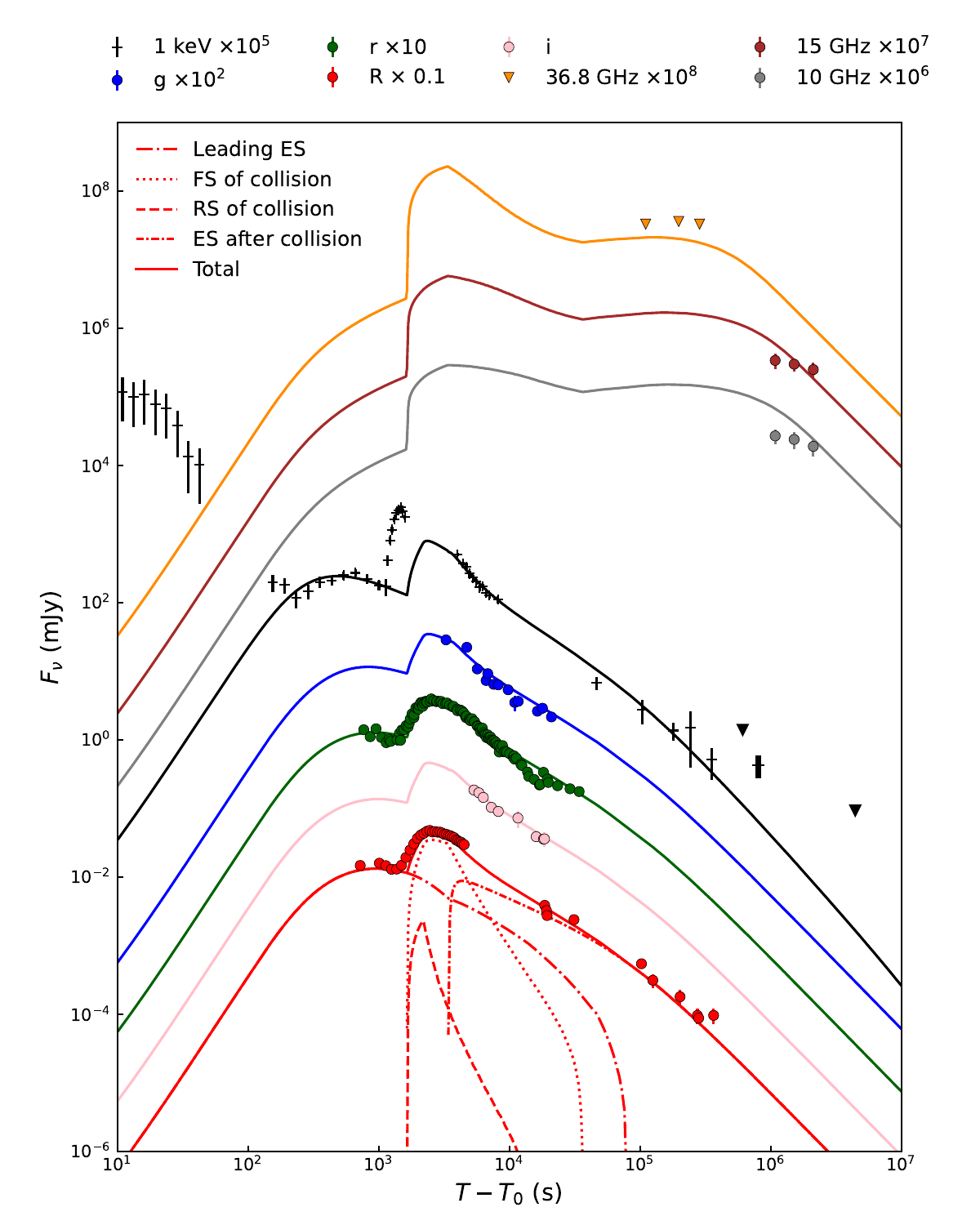}
	\caption{Modeling of afterglows within the scenario of the violent collision of two shells. The dash-dotted line is the emission of the leading external shock (marked as ES) propagating in the circumburst environment. The dotted and the dashed lines represent emission from the FS and the RS during the collision, respectively. The external shock after the FS crossing the leading shell is marked as the densely dashed-dotted line. The total flux from all emission components is given by the solid line for each band. \label{fig:model_collision}}
\end{figure}

We conduct numerical calculations to show that the shell collision scenario can well explain the multi-wavelength afterglow.
During the period of P2 (before the collision), the dynamics of the external shock of the first outflow is described by a generic model based on energy conservation \citep{Huang99,Peer12}. During the collision period (P3), a pair of shocks, including an FS and an RS, propagates into the leading shell and the rear shell, respectively. The dynamics of the FS/RS system is described by the mechanical model \citep{Zhang2002, Beloborodov06, Geng25b}. After the FS crosses the entire leading shell, a new external shock (ES) forms and propagates into the circumburst environment, producing observed emissions at late stages.
We calculate the time-evolving electron spectrum of each shocked region by solving the continuity equation in energy space, from which we compute the synchrotron and synchrotron self-Compton emission \citep{Geng18}.
In principle, there are more than ten free parameters for the two shells, and exploring the full parameter space is time-consuming. We adopt several parameters after some trials around typical values, i.e., the isotropic kinetic energy of the leading shell ($E_{\rm k,iso,1} = 2 \times 10^{52}$~erg) and the rear shell ($E_{\rm k,iso,2} = 10^{53}$~erg), $\Delta T \simeq 1000$~s (in the observer frame), $\Gamma_{1,0} = \Gamma_{2,0} = 80$, and circumburst density $n_{\rm ISM} = 1$~cm$^{-3}$. Here, we assume that the energy of the leading shell is comparable to the prompt emission energy detected by the WXT, which simultaneously alleviates the issue of excessive energy in the rear shell. We then perform a standard Bayesian approach to derive the reasonable values for the parameters left for the FS/RS (with subscript ``FR'') and ES (with subscript ``E''), i.e., the half jet opening angle $\theta_{\rm j}$, the equipartition parameters for post-shock electrons ($\xi_{\rm e,FR}$, $\xi_{\rm e,E}$) and magnetic field ($\xi_{\mathrm{B,FR}}$, $\xi_{\mathrm{B,E}}$), and indices of the shock accelerated electron spectrum ($p_{\rm FR}$, $p_{\rm E}$). As shown in Figure~\ref{fig:model_collision}, the multi-wavelength afterglow could be well explained with a set of parameters constrained (see results in Table \ref{tab:results_vc} and posterior distribution in Figure~\ref{fig:corner}) by the Markov Chain Monte Carlo method \citep{Foreman-Mackey13}.

We note that the isotropic kinetic energy of the rear shell adopted above, $E_{\rm k,iso,2}\sim10^{53}$~erg, appears to exceed the prompt isotropic energy $E_{\rm iso}\sim2\times10^{51}$~erg inferred from the EP-WXT fluence in the 0.5--4~keV band by a factor of $\sim$50. 

These two quantities are, however, fundamentally different physical measures. The EP-WXT band is narrow and likely samples only the soft tail of the prompt spectrum; the quoted $E_{\rm iso,1}$ may therefore represent only a lower limit to the bolometric isotropic energy. A proper $k$-correction to the standard 1--$10^{4}$~keV band could significantly raise the inferred prompt energy to $E_{\gamma,\rm iso}\sim10^{52}$~erg \citep{1993ApJ...413..281B, 2002A&A...390...81A}, bringing the actual ratio $E_{\rm k,iso,2}/E_{\gamma,\rm iso}$ to within a factor of $\sim$10. In the standard internal-shock model, only a small fraction ($\eta \sim 1$--$10\%$) of the bulk kinetic energy is converted into radiation \citep{1999ApJ...523L.113K, 2007ApJ...655..973K}, meaning the kinetic energy reservoir can legitimately be an order of magnitude larger than the observed gamma-ray output. Thus, $E_{\rm k,iso,2}\sim10^{53}$~erg is not extreme in the context of GRB energetics. If the late shell is collimated with a jet opening angle $\theta_j \sim 5^{\circ}$ (as constrained by our modeling), the true energy is reduced by a beaming factor $f_b \sim 0.004$, yielding a beaming-corrected kinetic energy of $\sim 4\times10^{50}$~erg.

We emphasize that this factor of five compares the kinetic-energy reservoirs of two distinct engine episodes, rather than their radiative outputs. Such non-monotonic energy release has observational precedent. In GRB 050502B, the giant X-ray flare had a fluence exceeding that of the nominal prompt emission and was interpreted as a second central-engine episode whose slower ejecta later refreshed the afterglow shock \citep{Falcone2006}. In GRB 231117A, the violent-collision model requires an injection shell that is more energetic than the impulsive shell, with a post-injection energy approximately 18 times the initial value \citep{Anderson2025}. For the long GRB 100418A, the late-time kinetic energy inferred under the energy-injection interpretation is at least $10^{2}$ times the prompt isotropic $\gamma$-ray energy \citep{Marshall2011}. Thus, $E_{\mathrm{k,iso},2}/E_{\mathrm{k,iso},1}\simeq5$ is within the range inferred for late energy-injection episodes. In EP250302a, the second episode can be powered by delayed fallback accretion onto the newly formed black hole. Its low radiative efficiency, $\eta_{\rm X}\simeq5.7\times10^{-3}$ (Section 4.4), possibly due to heavier baryon loading of the late ejecta \citep{Lei2017}, allows the shell to retain most of its kinetic energy while producing a much smaller soft-X-ray radiative output.

In a more general consideration, if the optical bump were attributed to the afterglow emission from a second independent relativistic jet, its optical peak flux being higher than that of the first main jet would require $E_{\rm k,iso,2} > E_{\rm k,iso,1} > E_{\gamma,\rm iso}$. This requirement is also consistent with the parameters derived from the two-component jet modeling ($E_{\rm k,iso,1}=5.0\times10^{52}$~erg and $E_{\rm k,iso,2}=1.4\times10^{53}$~erg; see Section~\ref{sec:alternatives}). We therefore suggest that the radiation efficiency of the X-ray flare may be smaller than that of the prompt emission phase, possibly due to a higher baryon loading in the jet launched by the late accretion process \citep{Lei2017}.

\section{Discussions}\label{discon}

\subsection{The Nature of EP250302a}
Despite the lack of a simultaneous gamma-ray detection, EP250302a exhibits numerous observational hallmarks characteristic of LGRBs. Its prompt emission duration, $T_{90} \approx 42$\,s, significantly exceeds the canonical 2 s threshold separating short and long GRBs. Although the limited energy band of the WXT precludes a direct measurement of the peak energy ($E_{\rm p}$) through spectral fitting, we explored a range of assumed $E_{\rm p}$ values to calculate the corresponding isotropic-equivalent energy ($E_{\mathrm{iso}}$). The resulting trajectory in the Amati relation ($E_{\rm p} \text{--} E_{\mathrm{iso}}$ plane) consistently falls within the LGRB population (Figure \ref{amati}d). Furthermore, the late-time multi-wavelength evolution—characterized by prominent X-ray flares and power-law afterglow decay—is similar to standard GRB phenomenology. Consequently, EP250302a probably originated from an LGRB progenitor, though gamma‑ray emission is undetected.

\subsection{Comparison with Shell Collision Candidates}

Many GRBs exhibit multi-peaked prompt emission in the Swift/BAT band, reflecting the intermittent ejection of multiple relativistic shells by the central engine \citep{1994MNRAS.269L..41R, 1997MNRAS.287..110S, 1997ApJ...490...92K, 2005SSRv..120..143B}. During the prompt phase, these shells undergo internal collisions and coalesce, generating the early $\gamma$-ray peaks, before transitioning into smoothly decaying standard afterglows observed in the Swift/XRT light curves. Theoretically, if the central engine reactivates hundreds to thousands of seconds post-burst and ejects a late-time fast shell, this shell can overtake the decelerated initial blast wave, triggering a violent shell collision. Such a process is expected to produce internal dissipation (manifested as a $\gamma$-ray/X-ray flare) and external shock refreshment (resulting in an achromatic multi-band rebrightening) with a distinct temporal delay. Logically compelling as it is, observational evidence that simultaneously satisfies all these criteria remains exceptionally scarce.

\begin{figure}[htbp]
	\centering
	\plotone{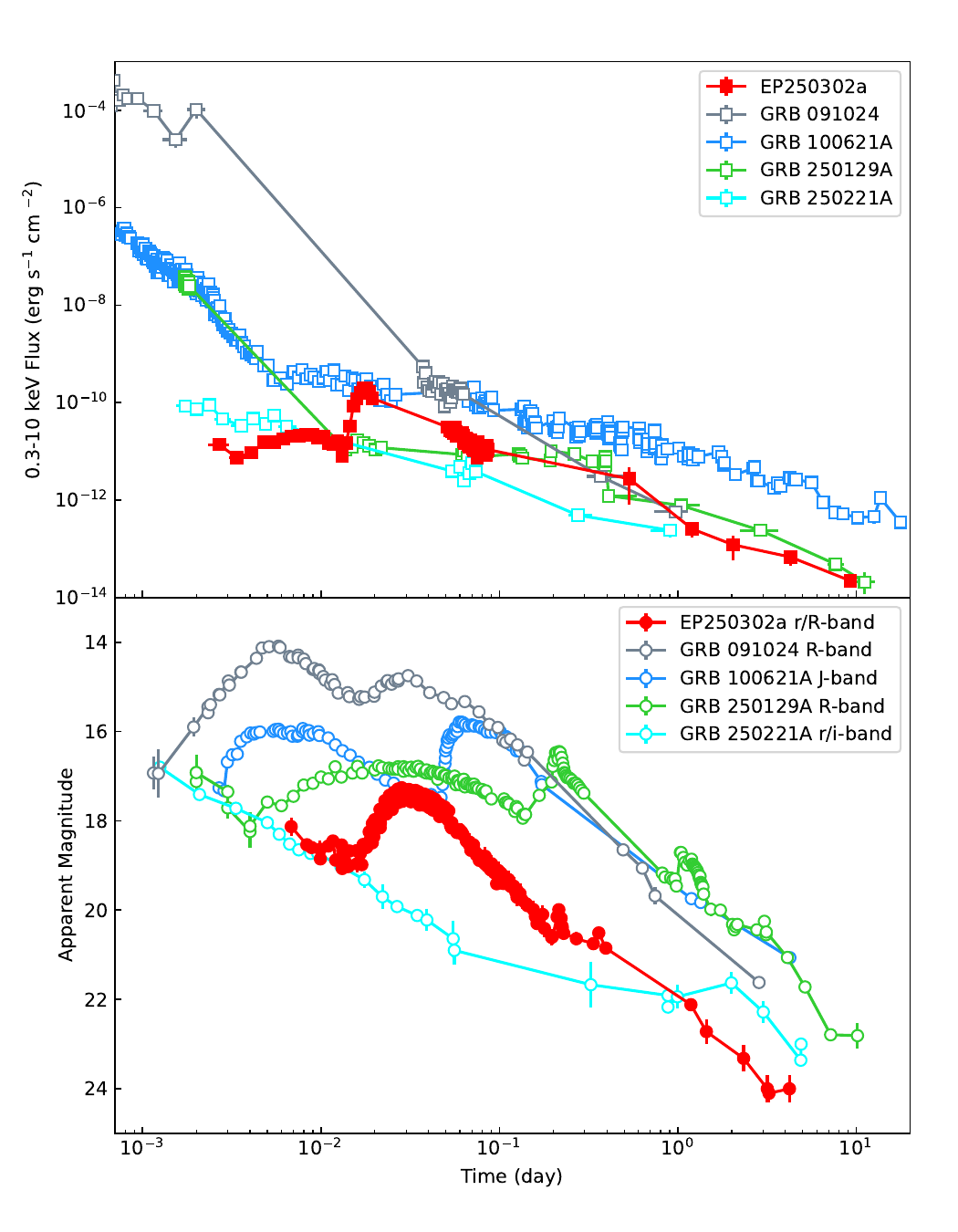}
	\caption{Multiwavelength light curves of a sample of GRBs with rebrightenings interpreted as violent shell collisions. Upper panel: 0.3-10~keV X-ray light curves for the comparison sample from \textit{Swift}/BAT-XRT and for EP250302a from the EP/FXT; lower panel: the densely sampled optical/near-infrared light curves. 
    \label{compare}}
\end{figure}

Figure \ref{compare} compiles several GRB events that have been previously interpreted within the framework of shell collisions \citep{2013ApJ...778...54V,2013A&A...560A..70G,2026A&A...710A.211A,2026MNRAS.tmp..179A}. Although these sources exhibit significant optical rebrightening or multi-peaked structures, a joint analysis with Swift/BAT data reveals that their optical peaks likely correspond to the early prompt emission phase. This correlation reflects successive shell ejections during the initial burst interval rather than late-time central engine reactivation. Unlike the distinct ``needle-like" X-ray flare captured in EP250302a (a flux increase by a factor of 25 within 400~s), none of these objects display such a feature in their subsequent Swift/XRT light curves. The ultra-long GRB 091024, characterized by a prompt emission duration exceeding $1000$ s, exhibited three distinct $\gamma$-ray peaks attributed to three individual ejected shells, which subsequently drove two optical rebrightening episodes at a few thousand seconds after the trigger \citep{2014MNRAS.445.1625N}. However, the comparatively sparse X-ray coverage during these rebrightening phases provided limited constraints on the shell-collision model.

EP250302a presents a similar phenomenology, but with the internal emission manifested in the X-ray band during the afterglow phase, where the temporal coverage during the brightening phase is more complete. Following the cessation of its prompt emission ($T_{90} \approx 42$~s), the early afterglow initiates a monotonic decay (P2, $\alpha_{X,d_1} \approx -1.0$). During the afterglow phase ($\sim 1.1$~ks), a luminous X-ray flare abruptly emerges (P3). Its exceptionally steep rise and decay ($\alpha_{X,r_2} \sim 15$, $\alpha_{X,d_3} \sim -15$) and narrow temporal width ($\Delta T/T \sim 0.5$) defy an external shock interpretation, clearly pointing to a late-time internal dissipation origin. Crucially, the optical rebrightening commences $\sim 200$~s after the onset of the X-ray flare ($\sim 1.3$~ks) and peaks at $\sim 2.5$~ks, consistent with the expected time lag within the ``internal flare to refreshed external shock'' framework. Furthermore, the optical-to-X-ray spectrum during the rebrightening phase is well-characterized by a single power law with an index of $\beta_{OX} = -0.7$, further supporting a common physical origin within the refreshed forward shock driven by the shell collision.

Benefiting from the wide-field soft X-ray monitoring and rapid alerting capabilities of the EP, we have acquired a comprehensive multiwavelength dataset for this event, spanning continuously from the prompt emission through the late-time flare to the subsequent optical rebrightening. This well-sampled temporal coverage enables us to quantitatively constrain the properties of the rear shell, including its launch time ($\Delta T \sim 1000$~s), isotropic kinetic energy ($E_{\rm k,iso,2} \sim 10^{53}$~erg), and the ratio of the bulk Lorentz factors ($\Gamma_{2,0}/\Gamma_{1,0} = 0.98$--$2.27$). Consequently, these parameters allow us to integrate late-time central engine reactivation, internal dissipation, and external shock refreshment into a single framework. EP250302a provides a clear, observationally resolved link between a late X-ray flare and optical rebrightening, yielding a robust observational basis for the detailed quantitative modeling of violent shell collisions.

\subsection{Alternative models}\label{sec:alternatives}
Numerous studies have invoked a two-component jet scenario to interpret the multi-wavelength rebrightening features observed in GRB afterglows. We initially adopted this two-component model to fit our multi-band data. In this model, the early optical afterglows and the interval P2 X-ray bump are dominated by the slow wide jet. The optical bump and the P4 X-ray light curve are dominated by the off-axis emission from the fast narrow jet component. The late afterglows (P5) are again dominated by the slow wide one. Such a model predicts achromatic behavior in afterglows, which fails to reproduce the chromaticity of the optical and X-ray bumps. In particular, the fast rise ($\alpha_{\rm X,r_2} \sim 15$) and sharp decay ($\alpha_{\rm X,d_3} \sim -15$) in interval P3 of the X-ray light curve (``needle-like'', see Figure \ref{EP250302a_lc}) are difficult to interpret within the framework of an external shock. A more natural explanation is the internal emission from a rear shell ejected during a reactivation of the central engine.

We therefore also consider a collisionless two-shell model (see Figure \ref{2comp_lc}). The onset time of the rear shell (produced by central engine reactivation), i.e., $\sim T_0 + 1100$\,s, is delayed relative to the initial trigger (leading shell). In this scenario, the P3 X-ray flare is interpreted as internal emission from the rear shell. The early optical afterglows and the interval P2 X-ray bump are dominated by the external emission from the leading shell. The optical bump and the P4 X-ray light curve are dominated by the external emission from the rear shell. The late afterglows are again dominated by the leading one.





To test these hypotheses, we utilized the \texttt{PyFRS}\footnote{\url{https://github.com/leiwh/PyFRS}} code \citep{Lei2016, 2024ApJ...974..221F} to construct a two-component jet and a collisionless two-shell model. For the collisionless two-shell configuration, the launch time of the rear shell was set at $T_0+1100$ s (in the observer frame). For simplicity, we assumed a top-hat jet profile propagating into an ISM environment and adopted an on-axis viewing geometry ($\theta_{\text{obs}} = 0$). Parameter space exploration was performed via Markov Chain Monte Carlo (MCMC) simulations using the \texttt{emcee} package \citep{2013PASP..125..306F}. We employed 30 walkers for 20000 steps, discarding the initial 50\% as the burn-in phase. The resulting best-fit light curves are presented in Figure \ref{2comp_lc}. 

Under typical afterglow parameters, the post-deceleration decay index typically settles near $\alpha \sim -1.0$. A transition to a steeper slope of $\alpha \sim -2.0$ is expected only after a jet break occurs. Our second component exhibits a post-peak decay index of $\alpha_{\rm o,d2} \sim -1.8$, suggesting that a jet break may have occurred shortly after the peak. Based on our fitting results for the collisionless two-shell model, this corresponds to a relatively narrow jet opening angle of $\theta_{\rm j,2nd} = 0.79^\circ$. For an ISM environment, the jet-break time is 
\begin{equation}
    t_{\rm break} = 5.8 \times 10^3 \text{ s } \hat{z} E_{52}^{1/3} \theta_{\rm j,-1}^{8/3} n_0^{-1/3},
\end{equation}
where $\hat{z}=(1+z)/2$ is the redshift correction factor. We estimate that the jet break occurs at $\sim T_0 + 1700$ s.

However, the collisionless two-shell model faces several physical challenges. In principle, the rear shell should propagate through the low-density tail evacuated by the leading blast wave. According to the Blandford--McKee solution, the ambient density encountered by the rear shell should be orders of magnitude lower than that encountered by the leading shell. In contrast, the best-fit parameters require nearly identical densities for the two components. Reducing the density encountered by the rear shell would require extreme values of the total kinetic energy or electron energy fraction to maintain the observed flux level. Furthermore, based on the current fit, the rear shell would inevitably catch up with the decelerating leading shell. Using Equation \ref{eq.tcoll}, we estimate that such a collision would occur at $t_{\rm coll}\approx T_0+6100$\,s in the observer frame.

We quantitatively compared the violent-collision model with these two alternatives using the Bayesian Information Criterion (BIC). All models were fitted to the same set of 202 multi-band measurements with the same treatment of upper limits. The BIC is defined as ${\rm BIC}=\chi^2+k\ln N$, where $\chi^2$ is the best-fit chi-square, $k$ is the number of free parameters, and $N$ is the number of measurements. The results are summarized in Table \ref{tab:model_comparison}.

\begin{deluxetable}{lcccc}
\tablecaption{Quantitative comparison of the three afterglow models based on the Bayesian Information Criterion.\label{tab:model_comparison}}
\tablewidth{0pt}
\tablehead{
\colhead{Model} &
\colhead{$N$} &
\colhead{$k$} &
\colhead{BIC} &
\colhead{$\Delta\mathrm{BIC}$}
}
\startdata
Violent collision       & 202 & 7 & 1940.74 & 0      \\
Two-component jet       & 202 & 14 & 2012.72 & 71.98  \\
Collisionless two-shell & 202 & 15 & 2261.22 & 320.48 \\
\enddata
\tablecomments{
$N$ is the number of measurements and $k$ is the number of free parameters. The Bayesian Information Criterion is calculated as $\mathrm{BIC}=\chi^2+k\ln N$, where $\chi^2$ is the best-fit chi-square. $\Delta\mathrm{BIC}$ is measured relative to the violent-collision model. All models are fitted to the same multi-band data set with the same treatment of upper limits.}
\end{deluxetable}

The violent-collision model yields the lowest BIC value, ${\rm BIC}=1940.74$, compared with ${\rm BIC}=2012.72$ for the two-component-jet model and ${\rm BIC}=2261.22$ for the collisionless two-shell model. Relative to the violent-collision model, the corresponding differences are $\Delta{\rm BIC}=71.98$ and $\Delta{\rm BIC}=320.48$, respectively. This statistical preference is obtained with the fewest free parameters among the three models. Therefore, within the set of models considered here, the BIC comparison formally favors the violent-collision scenario.

This quantitative preference also complements the physical motivation for the violent-collision model. Unlike the two-component-jet model, which describes the emission as a superposition of two independently prescribed jet components, the violent-collision model causally connects the narrow P3 X-ray flare and the delayed achromatic rebrightening as successive manifestations of the same late shell. The flare arises from internal dissipation of the late ejecta, while the rebrightening is produced when the shell collides with the decelerated leading blast wave.

\subsection{Central-engine reactivation}\label{sup.CE}
The chromatic behavior in the early afterglow of EFXT EP250302a enables us to investigate the nature of its central engine. The fast rise and sharp decay in the X-ray flare P3, as well as its temporal width-to-arrival time ratio of $\Delta T/T \sim 0.5$, pose a challenge to the standard external shock model, suggesting that the flare instead comes from a second shell launched at a late time. The physical origin of this rear shell is unclear. A possibility is the late-time reactivation of the central engine \citep{2007ApJ...671.1903C}.

There are two popular central engine models for powering an extremly relativistic jet (like GRBs), i.e., the black-hole model and the magnetar model. The maximal total rotation energy of the millisecond magnetar is 
\begin{eqnarray}
    E_{\rm rot} =\frac{1}{2} I \Omega^2 \simeq 2\times 10^{52} {\rm erg} M_{1.4} R_6^2 P_{0,-3}^{-2},
    \label{eq:Emagnetar}
\end{eqnarray}
where $I$ is the moment of inertia, $\Omega$ is the initial angular frequency of the magnetar, and $M_{1.4}=M/1.4M_\odot$. For EP250302a, the isotropic kinetic energy of the rear shell is $E_{\rm k,iso,2} = 10^{53}$~erg (see Section \ref{viocoll}), which is significantly larger than the maximal total energy available from a magnetar (assuming isotropic emission from the magnetar). This tension could be alleviated if the magnetar emission is collimated. We therefore favor a black-hole central engine with fallback accretion as the origin of the late-time activity. Two mechanisms are considered to power the relativistic jet in a GRB-like event for a black hole central engine: the neutrino–antineutrino annihilation mechanism, which liberates gravitational energy from the accretion disk \citep{PWF1999,Gu2006,Janiuk2007,Liu2015,Lei2017}, and the Blandford-Znajek \citep[hereafter BZ]{1977MNRAS.179..433B} mechanism, which extracts the spin energy of the Kerr black hole \citep{Lei2005,Liu2015,Lei2017}.

By integrating the X-ray flare flux, we derive an isotropic-equivalent energy of $E_{\rm X,iso} = 5.7 \times 10^{50}$ erg. The radiative efficiency of the flare is thus $\eta_{\rm X} = E_{\rm X,iso}/(E_{\rm k,iso}+E_{\rm X,iso})= 5.7\times 10^{-3}$. Using the fitting results of jet opening angle of $\theta_{\rm j} \approx 5^\circ$ (see Section \ref{viocoll}), we find $f_{\rm b} \approx 0.004$. Taking the flare duration of $\sim 1000$~s, we obtain an average luminosity for the rear shell of $ \bar{L}_{\rm jet,2} = E_{\rm iso,2} f_{\rm b}/(1000/(1+z))s \sim 8.6\times 10^{47} {\rm erg \ s^{-1}}$. Even for an optimistic total accreted mass of $\sim M_\odot$ during the X-ray flare phase (P3), the corresponding accretion rate would be $\sim10^{-3}M_\odot\ {\rm s^{-1}}$. For a black hole with this accretion rate, the neutrino–antineutrino annihilation power would be $\sim 2.8 \times 10^{42} \rm  erg  \ s^{-1}$ if we take a black hole mass of $3M_\odot$ and a spin $a_\bullet \sim 0.9$ \citep{Lei2017,Liu2017}. Therefore, the neutrino–antineutrino annihilation mechanism is not powerful enough to produce the emission of the second shell observed in EP250302a.


A more plausible scenario for the late-time activity in the black hole model is the BZ mechanism \citep{Wu2013}. In the fallback-accretion scenario, a later accretion episode can have a larger integrated energy budget than the prompt episode, even if its luminosity is lower. Material that remains bound after the initial explosion can circularize and accrete at a later time, launching a second energetic episode. If the ejecta produced in this episode are more heavily baryon loaded, only a small fraction of their kinetic energy needs to be dissipated as the observed soft-X-ray flare. The large kinetic reservoir of the rear shell and its relatively low radiative efficiency are therefore physically linked to the properties of a delayed, black-hole-powered accretion episode. The BZ jet power is \citep{Wu2013, Lei2017, 2024ApJ...974..221F,Lei2026}
\begin{eqnarray}\label{Eq.EB}
 \dot{E}_{\rm B} & = & 9.3 \times 10^{53} a_\bullet^2 \dot{m}  X(a_\bullet) \ {\rm erg \ s^{-1}},
 \label{eq:Ebz}
\end{eqnarray}
where $X(a_\bullet)=F(a_\bullet)/(1+\sqrt{1-a_\bullet^2} )^2$,  and $F(a_\bullet)=[(1+q^2)/q^2][(q+1/q) \arctan q-1]$. Here $q= a_{\bullet} /(1+\sqrt{1-a^2_{\bullet}})$. Taking a black hole spin of $a_\bullet \sim 0.9$, we estimate the average accretion rate of $\dot{M}_{\rm acc,2}\simeq 2.8 \times 10^{-6} M_\odot \ s^{-1}$ based on the average jet luminosity \citep{Wu2013}. This corresponds to a fallback mass of $M_{\rm fb,2} \approx 1.3 \times 10^{-3} M_{\odot}$, a value consistent with the mass predicted for fallback accretion in LGRB progenitors \citep[e.g.,][]{2010MNRAS.401.1465C, Wu2013}.

For comparison, we also estimate the black-hole accretion rate and accreted mass during the prompt episode by assuming the same central engine in operation. The leading shell has an isotropic kinetic energy of $E_{\rm k,iso,1}\sim2\times10^{52}$~erg and a duration of $T_{90}\sim42$~s in the observer frame. Assuming the same beaming factor as for the rear shell, its average power is $\bar{L}_{\rm jet,1}\simeq 4.5\times10^{48}\ {\rm erg\ s^{-1}}$. This corresponds to an average accretion rate of $\dot{M}_{\rm acc,1}\simeq1.5\times10^{-5}M_{\odot}\ {\rm s^{-1}}$. During the source-frame prompt duration of $T_{90}/(1+z)\sim20$~s, the total accreted mass is therefore $M_{\rm fb,1}\simeq2.9\times10^{-4}M_{\odot}$. Thus, although the instantaneous accretion rate during the prompt episode is about five times higher than that during the flare episode, its duration is approximately 24 times shorter. The total mass accreted during the late episode is consequently about five times larger than that accreted during the prompt episode. This may account for the larger kinetic energy of the rear shell without requiring a higher instantaneous central-engine power during the flare. Although this estimate assumes the same beaming factor for the two episodes, it demonstrates that the inferred energy ratio is physically plausible within the fallback-accretion scenario.

For the late fallback episode responsible for launching the rear shell, we estimate the characteristic fallback radius by assuming $M_\bullet=3M_{\odot}$ and adopting $t_{\rm fb}\sim1500/(1+z)$\,s, motivated by the observed delay of the X-ray flare:
\begin{eqnarray}
    R_{\rm fb}  \simeq 5.5\times 10^{10} {\rm cm} \left(\frac{M_\bullet}{3M_\odot}\right)^{1/3} \left(\frac{t_{\rm fb}}{700s}\right)^{2/3},
    \label{eq:Rfb}
\end{eqnarray}
which is highly reasonable for material stripped from the collapsar envelope \citep{Wu2013}.

However, as pointed out by \citet{Zhang2002}, a Poynting-flux-dominated injection may lead to a mild collision, which is usually very gradual. The abrupt signatures in the optical and X-ray bumps are expected from violent matter-dominated collisions. One possibility is that the magnetic energy in the rear shell is efficiently dissipated, via instabilities \citep[accelerating the jet]{Giannios2006} or the ICMART mechanism \citep[converting to radiation]{ZhangICMART2011}. Such internal dissipation may give rise to the significant X-ray flare (P3) observed in EP250302a.

\section{Conclusions}\label{con}

EP250302a represents a well-observed case of a violent two-shell collision. The comprehensive multi-wavelength dataset available for this source enables an in-depth investigation into the X-ray flare and the intricate details of shell-shell interactions. Notably, thanks to the rapid response and high sensitivity of the EP in the X-ray band, this event presents a distinct case where the clear signatures of a violent collision are captured simultaneously in both X-ray and optical bands. In particular, the distinct X-ray flare serves as a clear indicator of internal radiation, offering a valuable opportunity to diagnose the central engine’s reactivation. Alternative scenarios, such as the two-component jet model and the collisionless two-shell model, could potentially account for the observed multi-wavelength afterglow light curves; however, the two-component jet model fails to reproduce the ``needle-like'' X-ray flare. The collisionless two-shell model requires a significant ambient density, which is in conflict with the BM solution.

EP250302a exhibits observational characteristics closely matching those of classical GRBs. Despite the absence of a simultaneous gamma-ray trigger, its spectral parameters, the prominent X-ray flare driven by central-engine reactivation, and the late-time afterglow evolution are highly reminiscent of the LGRB population. Taking advantage of the high-sensitivity, wide-field focusing imaging provided by the WXT onboard the EP, we successfully captured the synchronized X-ray and optical signatures of this violent collision. Due to their relatively faint early-stage emission, such soft X-ray bursts provide an ideal ``low-background'' window for observing radiation associated with late-time central engine activity, which may be challenging to achieve in the Swift and Fermi eras. Our findings suggest that the EP mission will play an indispensable role in unveiling the physical origins of soft X-ray transients.

\begin{acknowledgments}
W.H. Lei acknowledges support by the National Natural Science Foundation of China under grants 12473012 and 12533005.
J.J. Geng acknowledges support by the National Natural Science Foundation of China under grants 12622309 and 12273113.
AW is supported by the National Natural Science Foundation of China under Grant Number 12503018 and the China Postdoctoral Science Foundation under Grant Numbers 2025M773197 and 2025T180874.
We acknowledge the support of the staff of the Nanshan station of the Xinjiang Astronomical Observatory (XAO) using the Photometric Auxiliary Telescope -17 (PAT-17). This work was partially supported by Tianshan Innovation Team Program of Xinjiang Uygur Autonomous Region, No. 2024D14015. This work was partly supported by the Urumqi Nanshan Astronomy and Deep Space Exploration Observation and Research Station of Xinjiang (XJYWZ2303). 
EA acknowledges support from the Science Committee of the Ministry of Science and Higher Education of the Republic of Kazakhstan (Grant No. AP26103591) and the Nazarbayev University Faculty Development Competitive Research Grant Program (no. 040225FD4713). TK acknowledges support from the Science Committee of the Ministry of Science and Higher Education of the Republic of Kazakhstan (Grant No. AP26102915) and the CAS-ANSO Fellowship, Grant No. CAS-ANSO-FA-2024-06. 
AP, NP, AV, AM thank the NKTRT committee for providing time for observing on the BTA-6m telescope. EK thanks the ANGARA project for use of the AZT-33IK telescope of Mondy observatory. 
A.S. acknowledges financial support from the Centre national d’études spatiales (CNES), France (ROR: \url{https://ror.org/04h1h0y33}) within the framework of the SVOM mission.
\end{acknowledgments}

\begin{contribution}
Weimin Yuan has been leading the Einstein Probe project as Principal Investigator since the mission proposal stage. Weihua Lei, Xiangyu Wang, Tao An, Dong Xu, Shaoyu Fu, Cuiyuan Dai, and Ailing Wang initiated this study. Weihua Lei, Xiangyu Wang, Dong Xu, Shaoyu Fu, and Xuefeng Wu coordinated the scientific investigation of the event and chaired the subsequent discussions. Weihua Lei, Jinjun Geng, and Tao An contributed to the theoretical research of the event. Weihua Lei, Xiangyu Wang, Dong Xu, Xuefeng Wu, He Gao, Tao An, Jinjun Geng, Yuanchuan Zou, Yunwei Yu, Rongfeng Shen, Binbin Zhang, Liangduan Liu, Ye Li, Yi-Han Wang, Bing Zhang, Valerio D'Elia, Ruben Salvaterra and Massimiliano De Pasquale participated in discussions and provided valuable suggestions.
Cuiyuan Dai and Xiangyu Wang planned and submitted the EP-FXT follow-up observations. Cuiyuan Dai, Shuaiqing Jiang, and Hongzhou Wu processed the light curves and spectral data from EP-WXT and FXT. Dong Xu, Zipei Zhu, Xing Gao, Abdusamatjan Iskandar, Shahidin Yaqup, Tuhong Zhong, Ali Esamdin, Chunhai Bai, and Yu Zhang acquired early optical data using HMT and PAT17. Junjie Jin acquired optical data using TNT. Alexei Pozanenko, Nicolai Pankov, Alexandr Moskvitin, Alina Volnova, Toktarkhan Komesh, Ernazar Abdikamalov, Dilda Berdikhan, and Zhanat Maksut acquired and processed optical photometric data using NUTTelA-TAO, Mondy/AZT-33IK, SAO-RAS/Zeiss-1000, and CrAO/ZTSh. D. B. Malesani, L. Izzo, R. A. J. Eyles-Ferris, A. Saccardi, B. Schneider, J. Palmerio, and N. R. Tanvir triggered the VLT/MUSE spectroscopic observations and performed the data reduction and spectral analysis.
Shaoyu Fu processed the Swift-UVOT photometric data and performed unified photometry on all optical photometric images. Zipei Zhu processed the spectral data taken by XL216. Luca Izzo processed the spectral data taken by VLT-MUSE. A. Volvach and L. Volvach acquired and processed early radio data using RT-22. Ailing Wang and Tao An acquired and processed late-time radio data using the VLA.
Shaoyu Fu performed temporal and dual-component forward shock model fitting for the multi-wavelength light curve data. Jinjun Geng modeled the multi-wavelength light curves using the violent collision model and wrote the relevant sections. Weihua Lei, Tao An, and Shaoyu Fu estimated the relevant physical quantities using the fallback accretion model. Shaoyu Fu, Cuiyuan Dai, and Ailing Wang co-drafted the manuscript with assistance from all authors.

\end{contribution}



\appendix

\section{Observation results}
\begin{figure*}[htbp]
	\centering
	\plotone{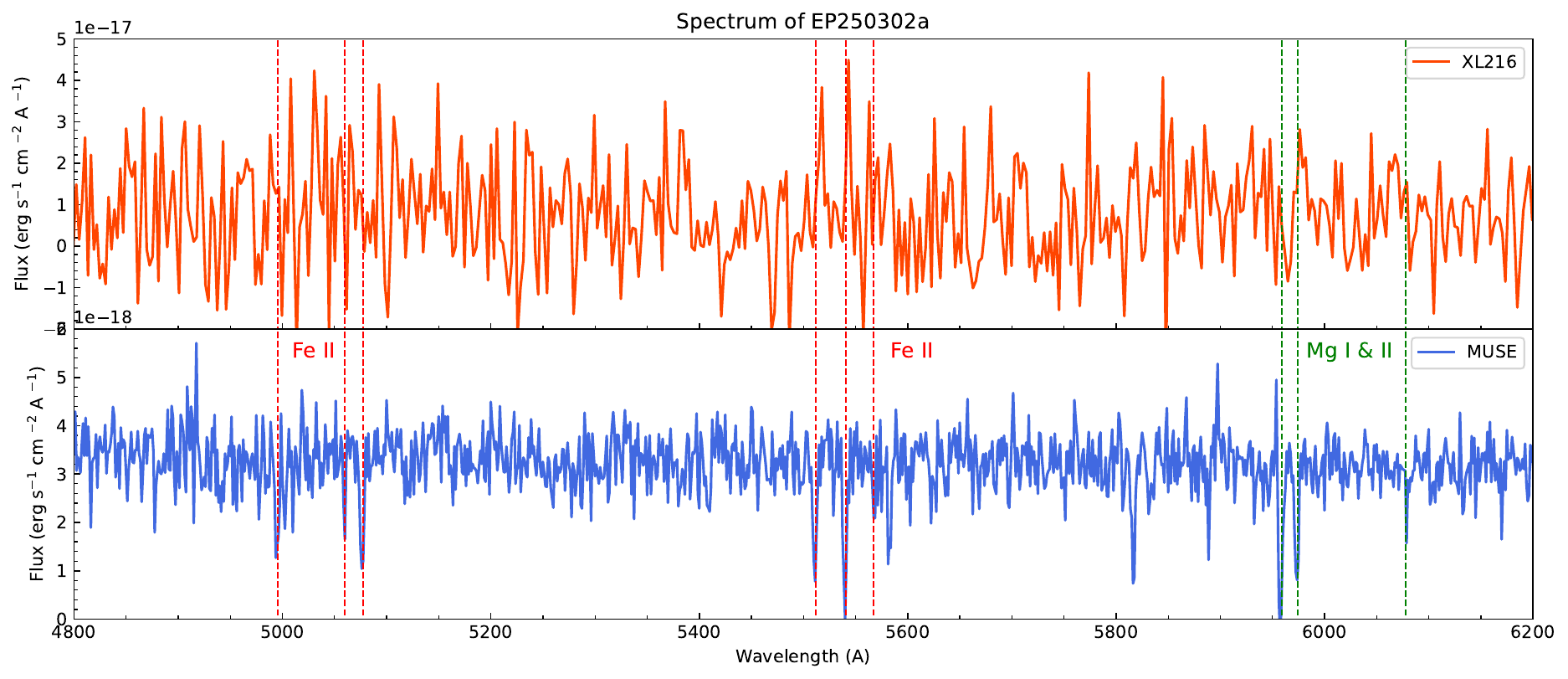}
	\caption{Optical spectra of EP250302a. The upper panel displays the spectrum obtained with the Xinglong 2.16m telescope, while the lower panel shows the VLT/MUSE spectrum. The red spectral region is heavily contaminated by sky emission lines and shows no obvious absorption lines from the target; therefore, we only present the 4800–6200 \AA portion of the spectrum here. Vertical dashed lines denote the identified absorption features in the MUSE data at $z = 1.131$. The redward portion of the MUSE spectrum is affected by sky emission lines (not shown in this figure, predominantly above 6200\,\AA).}
	\label{EP250302a_spec}
\end{figure*}

Figure \ref{EP250302a_spec} displays the spectra of EP250302a obtained with the Xinglong 2.16 m telescope and VLT/MUSE. Numerous metallic absorption lines are clearly visible in the plot, confirming a redshift of $z=1.131$. Table \ref{tab_FXT_obs} provides the observation log for EP-FXT, while Table \ref{tab_VLA_obs} details the VLA radio observation results. Table \ref{tab_phot} lists the photometry results across all optical bands. It should be noted that the magnitudes presented in the tables have not been corrected for Galactic extinction.

\begin{figure*}[htbp]
\gridline{\fig{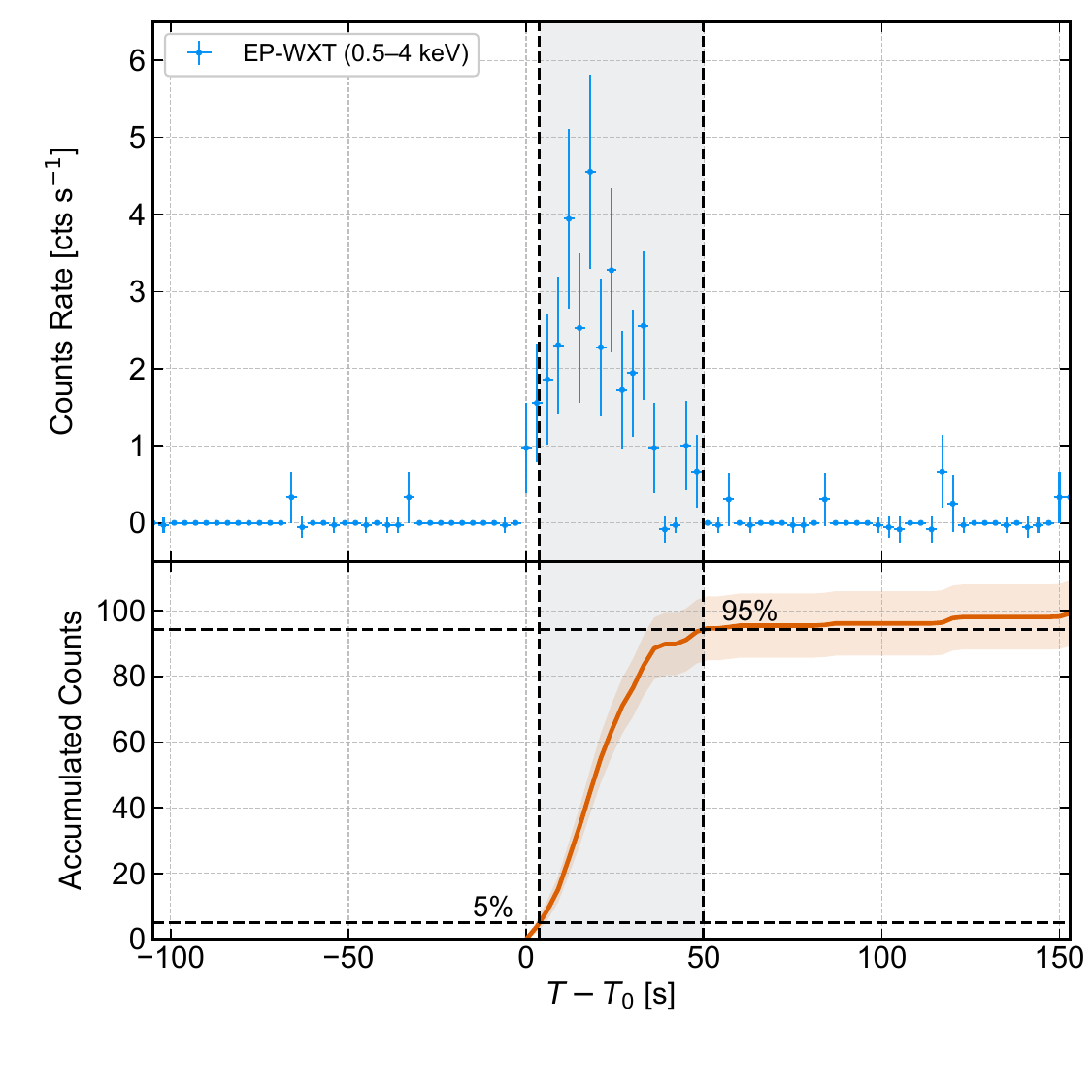}{0.46\textwidth}{(a)}
          \fig{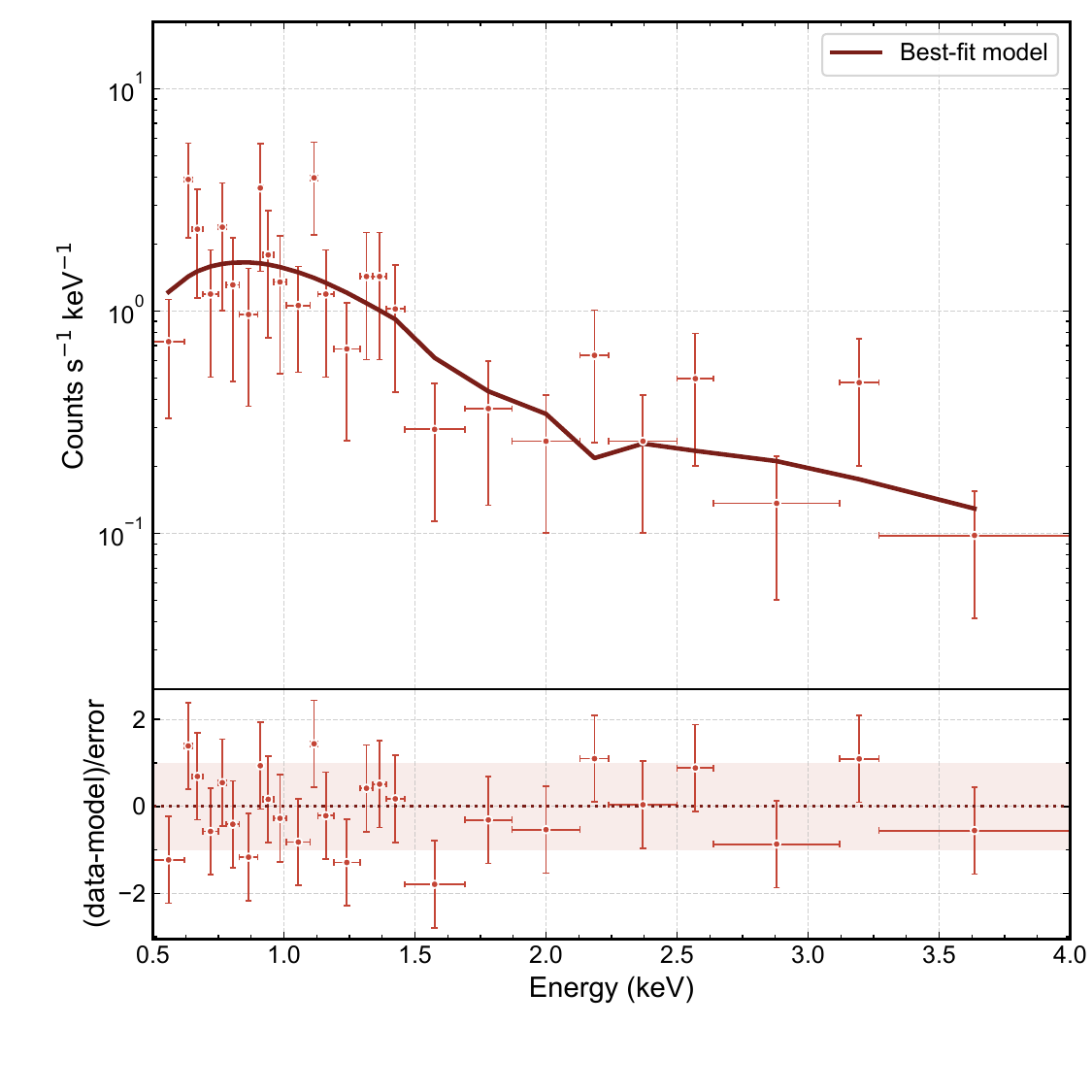}{0.46\textwidth}{(b)}
          }
\gridline{\fig{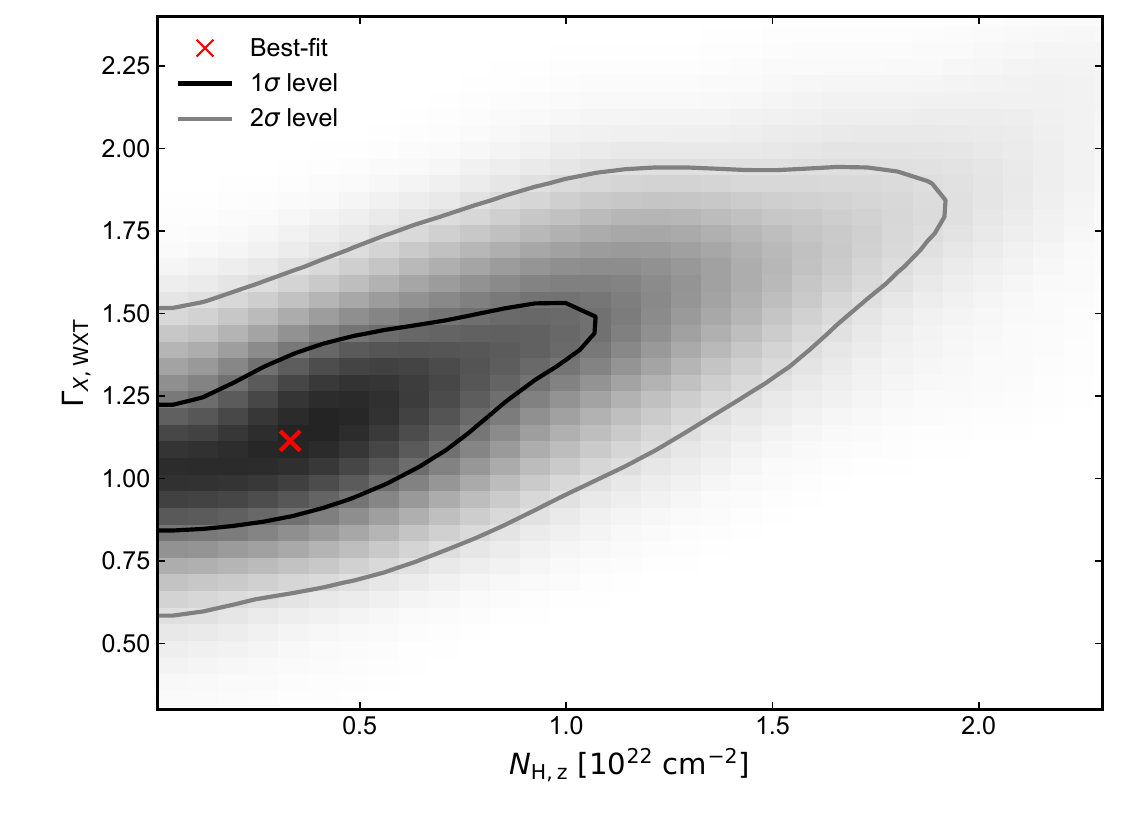}{0.46\textwidth}{(c)}
          \fig{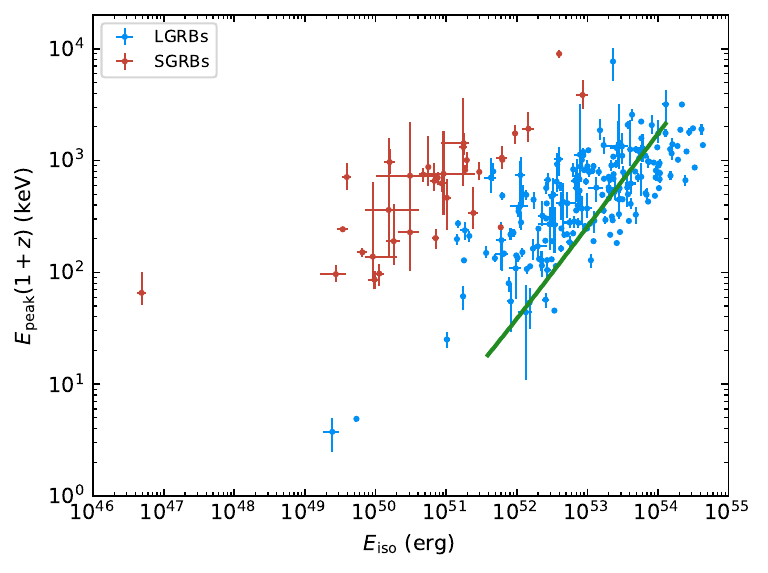}{0.46\textwidth}{(d)}
          }          
\caption{Prompt X-ray properties of EP250302a. 
(a) The upper panel displays the 0.5--4\,keV count rate of EP250302a observed with EP-WXT, while the lower panel shows the corresponding accumulated counts. The two vertical dashed lines indicate $T_{05} = 3 \, \rm s$ and $T_{95} = 45 \, \rm s$, which correspond to the epochs when the accumulated counts reach 5\% and 95\% of the total fluence, respectively. 
(b) The upper panel presents the EP-WXT spectrum, with the solid curve representing the best-fit absorbed power-law model. The lower panel shows the residuals relative to this best-fit model. 
(c) Two-dimensional confidence contours of the photon index $\Gamma_X$ and the intrinsic absorption column density $N_{\rm int}$. The black and blue contours denote the 1$\sigma$ and 2$\sigma$ confidence levels, respectively. 
(d) $E_{\rm p,i}-E_{\rm{iso}}$ diagram of EP250302a (green line). The green solid line is obtained by assuming a Band model with ($E_{\rm p}$) ranging from 4 keV to 1 MeV; as shown, the trajectory of EP250302a is consistent with the population of LGRBs. The comparison sample is adopted from \citet{2024ApJ...974..221F} and references therein.\label{Fig_WXT_ana}\label{amati}}
\end{figure*}

\begin{deluxetable*}{llccccc}[htbp]
\digitalasset
\tablewidth{0pt}
\tablecaption{Spectral results of X-ray observations of EP250302a. Errors represent the 1$\sigma$ uncertainties. The upper limit is quoted at the 95\% confidence level.}\label{tab_spectrum_fitting}
\tablehead{
\colhead{Instrument} & \colhead{Time Interval} & \colhead{Model} & \colhead{$\Gamma_X$} & \colhead{$N_{\rm int}$} & \colhead{Flux$^*$} & \colhead{CSTAT/(d.o.f.)} \\
\colhead{} & \colhead{} & \colhead{} & \colhead{} & \colhead{($10^{22} \, \rm{cm^{-2}}$)} & \colhead{($\rm erg\,cm^{-2}\,s^{-1}$)} & \colhead{}
}
\startdata
{EP-WXT} & [3\,s, 45\,s] & PL & $1.01_{-0.25}^{+0.42}$ & $<1.50$ & $5.39^{+0.90}_{-0.77} \times 10^{-9}$ & $35.16/37$ \\
\hline
\multirow{3}{*}{EP-FXT} & [154\,s, 240\,s] & PL & $2.60^{+0.44}_{-0.36}$ & $<0.93$ & $6.36^{+1.66}_{-1.07} \times 10^{-12}$ & $76.59/93$ \\
\cline{2-7}
 & [240\,s, 1.1\,ks] & PL & $2.37^{+0.14}_{-0.13}$ & $0.34^{+0.19}_{-0.18}$ & $1.13^{+0.05}_{-0.05} \times 10^{-11}$ & $404.81/474$ \\
\cline{2-7}
 & [1.1\,ks, 1.6\,ks] & PL & $1.97^{+0.05}_{-0.05}$ & $0.37^{+0.09}_{-0.09}$ & $9.60^{+0.24}_{-0.24} \times 10^{-11}$ & $763.53/811$ \\
\hline
EP-FXT & [4.4\,ks, 7.0\,ks] & PL & \multirow{3}{*}{$1.94^{+0.04}_{-0.04}$} & \multirow{3}{*}{$0.26^{+0.09}_{-0.09}$} & $1.45^{+0.04}_{-0.04} \times 10^{-11}$ & \multirow{3}{*}{$892.03/1008$} \\
\textit{Swift}/XRT & [3.7\,ks, 4.3\,ks] & PL &  &  & $3.97^{+0.20}_{-0.20} \times 10^{-11}$ &   \\
\textit{Swift}/XRT & [8.0\,ks, 8.4\,ks] & PL &  &  & $9.08^{+1.25}_{-1.15} \times 10^{-12}$ &   \\
\hline
EP-FXT & [44.7\,ks, 47.8\,ks] & PL & $1.99^{+0.27}_{-0.26}$ & \multirow{7}{*}{$0.26$ (fixed)} & $5.35^{+1.30}_{-0.97} \times 10^{-13}$ & \multirow{7}{*}{$418.51/457$} \\
EP-FXT & [1.19\,d, 1.22\,d] & PL & $1.76^{+0.47}_{-0.45}$ &  & $2.59^{+1.36}_{-0.80} \times 10^{-13}$ &   \\
EP-FXT & [1.99\,d, 2.15\,d] & PL & $1.98^{+0.40}_{-0.37}$ &  & $1.08^{+0.39}_{-0.26} \times 10^{-13}$ &   \\
EP-FXT & [3.85\,d, 4.29\,d] & PL & $2.46^{+0.85}_{-0.78}$ &  & $3.64^{+2.44}_{-1.19} \times 10^{-14}$ &   \\
EP-FXT & [9.19\,d, 9.75\,d] & PL & $3.43^{+1.38}_{-1.18}$ &  & $2.19^{+1.05}_{-0.69} \times 10^{-14}$ &   \\
\textit{Swift}/XRT & [1.39\,d, 4.22\,d] & PL & $1.84^{+1.31}_{-0.98}$ &  & $1.23^{+1.39}_{-0.69} \times 10^{-13}$ &   \\
\textit{Chandra} & [8.94\,d, 9.19\,d] & PL & $2.46^{+0.87}_{-0.79}$ &  & $2.28^{+1.26}_{-0.74} \times 10^{-14}$ &   \\
\enddata
\tablecomments{$^*$ The unabsorbed flux is measured in the $[0.5, 4.0]\,\rm{keV}$ energy range for EP-WXT and in the $[0.5, 10.0]\,\rm{keV}$ range for EP-FXT, \textit{Swift}/XRT, and \textit{Chandra}.}
\end{deluxetable*}


\begin{figure*}[htbp]
	\centering
	\plotone{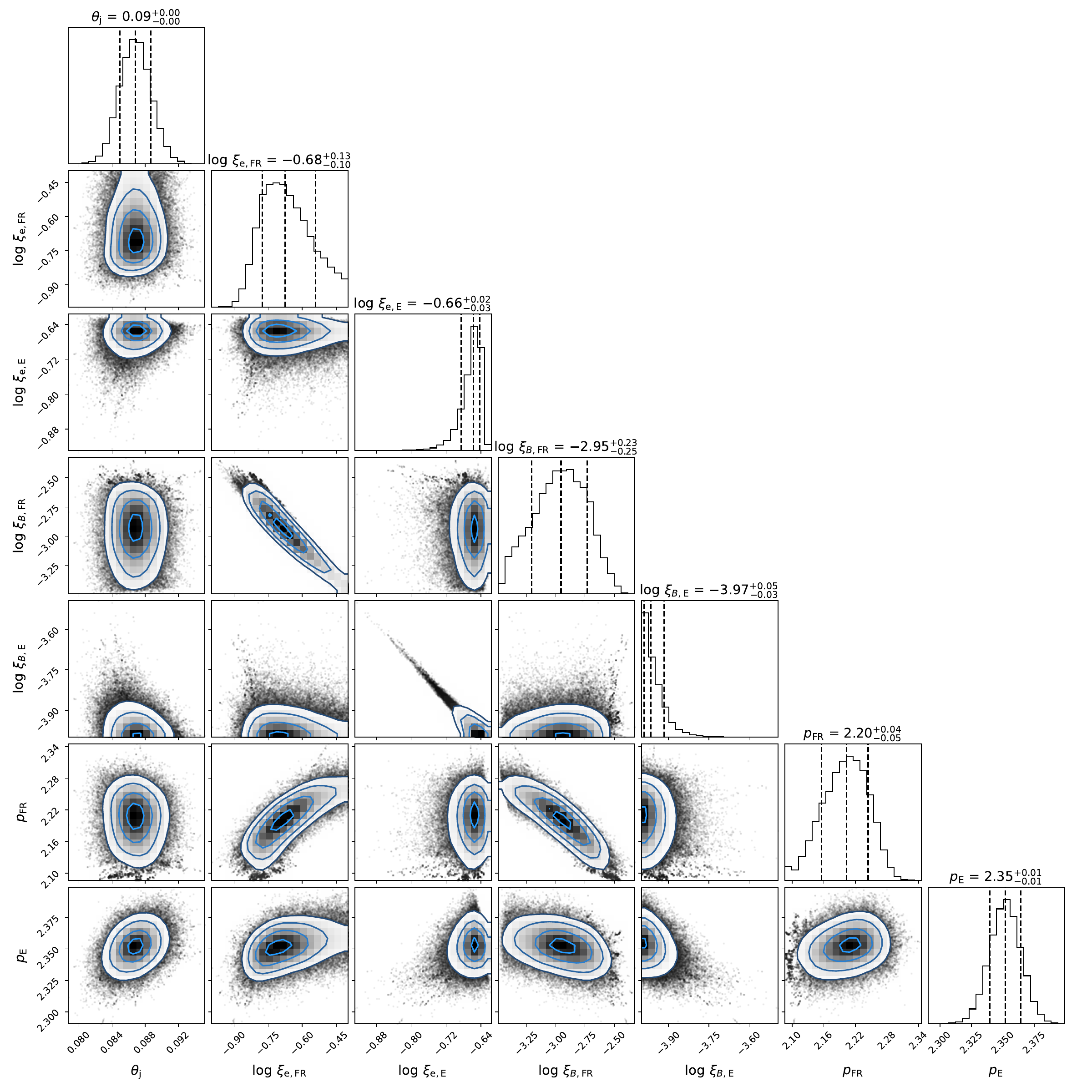}
	\caption{Violent collision model parameters constrained using MCMC. The corner plot shows one- and two-dimensional projections of the posterior probability distributions of seven parameters used in the model. The 1-dimensional histograms are marginal posterior distributions of these parameters. The vertical dashed lines indicate the 16th, 50th, and 84th percentiles of the samples, respectively, which are labeled on the top of each histogram. \label{fig:corner}}
\end{figure*}

\begin{figure*}[htbp]
\gridline{\fig{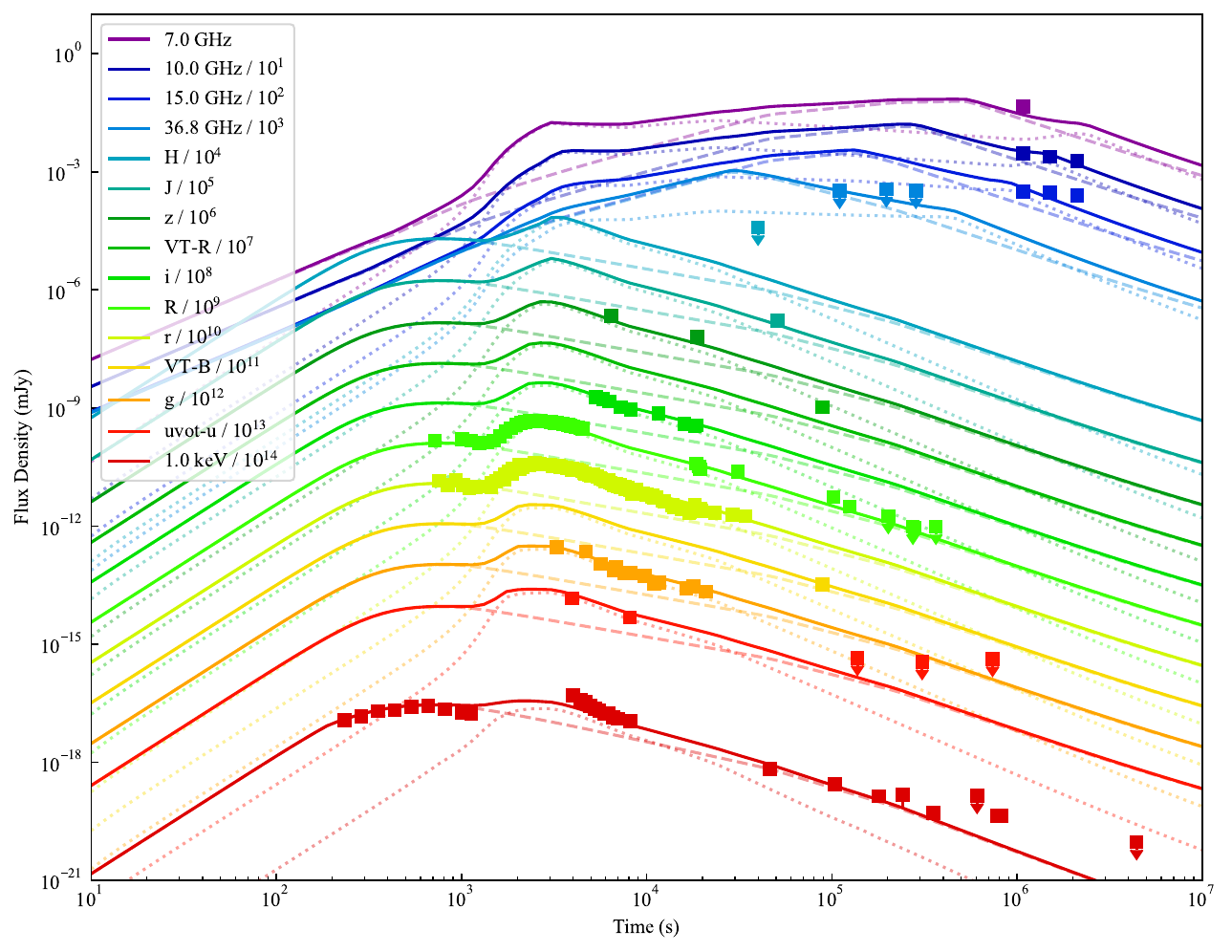}{0.46\textwidth}{(a)}
          \fig{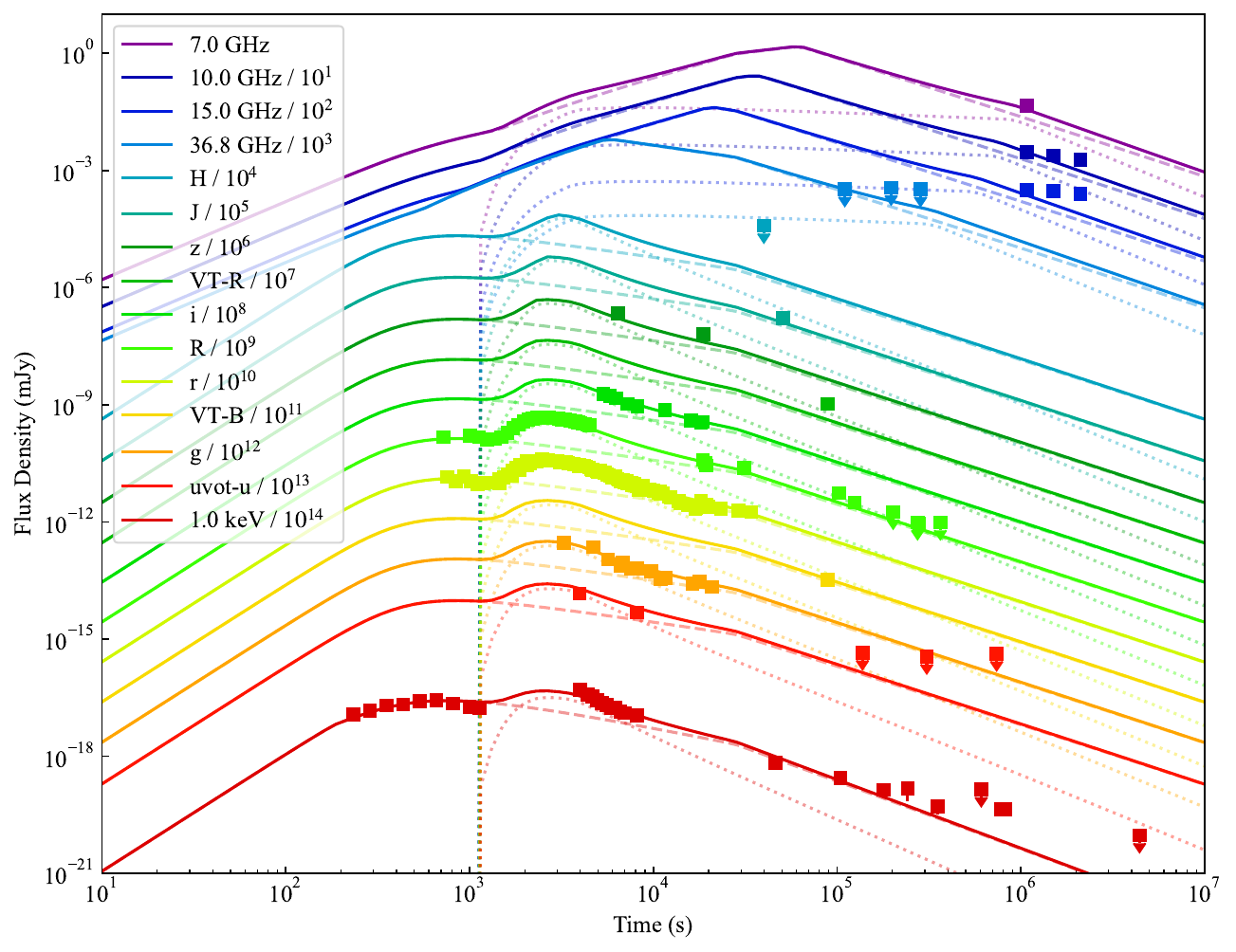}{0.46\textwidth}{(b)}
          }        
\caption{Modeling of afterglows with (a) two-component jet model and (b) collisionless two-shell model. (a) The wide jet is represented by dashed lines, and the narrow jet by dotted lines. The solid lines are the combination of the two components. (b) The primary jet is represented by dashed lines, and the second one by dotted lines. The solid lines are the combination of the contributions from the two shells. \label{2comp_lc}}
\end{figure*}


\begin{deluxetable*}{ccccc}[htbp]
\digitalasset
\tablewidth{0pt}
\tablecaption{Log of the EP-FXT follow-up observations.}\label{tab_FXT_obs}
\tablehead{
\colhead{ObsID} & \colhead{Start$\;$Time} & \colhead{End$\;$Time} & \colhead{Duration} & \colhead{Exposure}\\
\colhead{} & \colhead{(UTC)} & \colhead{(UTC)} & \colhead{(second)} & \colhead{(second)}}
\startdata
01709132186 & 2025-03-02T15:38:17 & 2025-03-02T17:39:12 & 7255 & 4499 \\
06800000453 & 2025-03-03T04:01:21 & 2025-03-03T04:51:32 & 3011 & 3011 \\
06800000457 & 2025-03-03T20:01:43 & 2025-03-03T20:52:00 & 3017 & 3017 \\
06800000459 & 2025-03-04T15:14:09 & 2025-03-04T19:16:38 & 14549 & 9073 \\
06800000464 & 2025-03-06T12:03:05 & 2025-03-06T14:29:46 & 8801 & 6078 \\
06800000465 & 2025-03-06T21:39:16 & 2025-03-06T22:29:58 & 3042 & 3042 \\
11900133760 & 2025-03-11T20:05:37 & 2025-03-12T09:30:05 & 48268 & 9030 \\
06800000542 & 2025-04-21T09:07:33 & 2025-04-21T11:33:13 & 8740 & 5829 \\
06800000543 & 2025-04-21T18:43:25 & 2025-04-21T19:33:07 & 2982 & 2982 \\
06800000556 & 2025-04-24T05:54:32 & 2025-04-24T08:13:25 & 8333 & 4994 \\
\enddata
\end{deluxetable*}


\begin{deluxetable*}{cccccccc}[htbp]
\digitalasset
\tablewidth{0pt}
\tablecaption{Log of the VLA follow-up observations.}\label{tab_VLA_obs}
\tablehead{
\colhead{Start$\;$Time} & 
\colhead{End$\;$Time} & 
\colhead{Frequency} & 
\colhead{BW} & 
\colhead{Beam Size} & 
\colhead{BPA} & 
\colhead{Peak Flux Density} & 
\colhead{rms} \\
\colhead{(UTC)} & 
\colhead{(UTC)} & 
\colhead{(GHz)} & 
\colhead{(GHz)} & 
\colhead{arcsec $\times$ arcsec} & 
\colhead{($^\circ$)} & 
\colhead{$\mu$Jy beam$^{-1}$} & 
\colhead{$\mu$Jy beam$^{-1}$}
}
\startdata
2025-03-15T02:17:50 & 2025-03-15T03:38:31 & 7  & 2 & 11 $\times$ 7.9  & $-$82 & 45 & 11  \\
2025-03-15T02:17:50 & 2025-03-15T03:38:31 & 10 & 4 & 11 $\times$ 6.9  & $-$75 & 27 & 6.8 \\
2025-03-15T02:17:50 & 2025-03-15T03:38:31 & 15 & 6 & 6.0 $\times$ 3.8 & $-$80 & 34 & 8.4 \\
2025-03-20T01:09:09 & 2025-03-20T02:30:45 & 10 & 4 & 19 $\times$ 4.5  & $-$74 & 24 & 6.5 \\
2025-03-20T01:09:09 & 2025-03-20T02:30:45 & 15 & 6 & 8.2 $\times$ 3.2 & $-$78 & 30 & 6.6 \\
2025-03-27T01:05:17 & 2025-03-27T02:26:52 & 10 & 4 & 9.0 $\times$ 5.6  & $-$77 & 19 & 5.6 \\
2025-03-27T01:05:17 & 2025-03-27T02:26:52 & 15 & 6 & 7.1 $\times$ 3.7  & $-$73 & 25 & 6.6 \\
\enddata
\tablecomments{
BW is the effective bandwidth of the final continuum image.
BPA is the synthesized beam position angle measured east of north.
}
\end{deluxetable*}

\begin{deluxetable*}{cccc}[htbp]
\digitalasset
\tablewidth{0pt}
\tablecaption{Fitting results of violent collision model \label{tab:results_vc}}
\tablehead{
\colhead{Parameters} & \colhead{Prior Type} & \colhead{Prior} & \colhead{Value} }
\startdata
$E_{\rm k,iso,1}$ ($10^{52}$ erg)  & fixed & -- & 2  \\
$E_{\rm k,iso,2}$  ($10^{52}$ erg) & fixed & -- & 10 \\
$\Gamma_{1,0}$ & fixed & -- & 80\\
$\Gamma_{2,0}$ & fixed & -- & 80\\
$n_{\rm ISM}$ (cm$^{-3}$) & fixed & -- & 1\\
$\theta_j$ (rad) & uniform & [0.05, 0.15] & $0.09_{-0.01}^{+0.01}$ \\
$p_{\rm FR}$  & uniform & [2.05, 2.8] & $2.20_{-0.05}^{+0.04}$\\
$p_{\rm E}$  & uniform & [2.05, 2.8] & $2.35_{-0.01}^{+0.01}$\\
$\xi_{\rm e,FR}$  & log-uniform & [$10^{-2}$, 0.5]  & $2.09_{-0.43}^{+0.73}\times10^{-1}$\\
$\xi_{\rm e,E}$  & log-uniform & [$10^{-2}$, 0.5]  & $2.19_{-0.15}^{+0.10}\times10^{-1}$\\
$\xi_{\mathrm{B,FR}}$  & log-uniform & [$10^{-4}$, $10^{-2}$] & $1.12_{-0.49}^{+0.78}\times10^{-3}$\\
$\xi_{\mathrm{B,E}}$  & log-uniform &  [$10^{-4}$, $10^{-2}$]  & $1.07_{-0.07}^{+0.13}\times10^{-4}$\\
\enddata
\end{deluxetable*}

\begin{deluxetable*}{cccc}
\tabletypesize{\scriptsize}
\tablecaption{Fitting results of alternative models}
\label{tb.AGmodel}
\tablewidth{0pt}
\tablehead{
\colhead{Parameters} & \colhead{Prior Type} & \colhead{Prior} & 
\colhead{Results}
}
\startdata
$E_{\rm k,iso,n}$ (erg) & log-uniform & $[10^{51}, 10^{54}]$ &  $5.01^{+0.74}_{-0.55}\times 10^{52}$  \\
$E_{\rm k,iso,w}$ (erg) & log-uniform & $[10^{52}, 10^{55}]$ &  $1.38^{+3.52}_{-0.36}\times 10^{53}$  \\
$\Gamma_{0,n}$ & log-uniform & $[10, 200]$ &  $70.8^{+6.8}_{-7.7}$ \\
$\Gamma_{0,w}$ & log-uniform & $[10, 500]$ &  $57.5^{+10.1}_{-0.51}$ \\
$p_n$ & uniform &  $[2.01, 3.0]$ &  $2.65^{+0.02}_{-0.04}$ \\
$p_w$ & uniform &  $[2.01, 3.0]$ &  $2.01^{+0.01}_{-0.01}$ \\
$n_{18}$ (cm$^{-3}$) & log-uniform& $[10^{-3}, 10^2]$ & $2.82^{+0.81}_{-1.04}\times 10^{1}$ \\
$\theta_{\rm obs,n}$ (degrees) & uniform & $[0.0, 45]$ & $4.55^{+0.21}_{-0.30}$ \\
$\theta_{j,n}$ (degrees) & uniform & $[0.1, 10]$ & $2.37^{+0.13}_{-0.17}$ \\
$\theta_{j,w}$ (degrees) & uniform & $[0.1, 20]$ & $6.97^{+0.55}_{-0.62}$ \\
$\epsilon_{e,n}$ & log-uniform & $[10^{-3}, 0.5]$ & $3.16^{+0.31}_{-0.34}\times 10^{-1}$ \\
$\epsilon_{e,w}$ & log-uniform & $[10^{-3}, 0.5]$ & $6.31^{+2.60}_{-4.86}\times 10^{-2}$ \\
$\epsilon_{B,n}$ & log-uniform & $[10^{-4}, 0.5]$ & $1.17^{+0.37}_{-0.13}\times 10^{-4}$ \\
$\epsilon_{B,w}$ & log-uniform & $[10^{-4}, 0.5]$ & $2.24^{+0.71}_{-0.76}\times 10^{-4}$ \\
\hline
$E_{\rm k,iso,1}$ (erg) & log-uniform & $[10^{51}, 10^{54}]$ &  $4.47^{+3.12}_{-2.85}\times 10^{54}$  \\
$E_{\rm k,iso,2}$ (erg) & log-uniform & $[10^{52}, 10^{55}]$ &  $2.57^{+1.79}_{-0.95}\times 10^{53}$  \\
$\Gamma_{0,1}$ & log-uniform & $[10, 200]$ &  $257^{+45.0}_{-52.9}$ \\
$\Gamma_{0,2}$ & log-uniform & $[10, 500]$ &  $120^{+21.0}_{-26.9}$ \\
$p_1$ & uniform &  $[2.01, 3.0]$ &  $2.01^{+0.01}_{-0.01}$ \\
$p_2$ & uniform &  $[2.01, 3.0]$ &  $2.46^{+0.03}_{-0.03}$ \\
$n_{18,1}$ (cm$^{-3}$) & log-uniform& $[10^{-3}, 10^1]$ & $3.09^{+16.4}_{-1.77}\times 10^{-3}$ \\
$n_{18,2}$ (cm$^{-3}$) & log-uniform& $[10^{-3}, 10^1]$ & $4.07^{+14.1}_{-2.45}\times 10^{-3}$ \\
$\theta_{\rm obs}$ (degrees) & uniform & fixed & 0 \\
$\theta_{j,1}$ (degrees) & uniform & $[0.1, 10]$ & $1.42^{+0.41}_{-0.20}$ \\
$\theta_{j,2}$ (degrees) & uniform & $[0.1, 20]$ & $0.79^{+0.23}_{-0.12}$ \\
$\epsilon_{e,1}$ & log-uniform & $[10^{-3}, 0.5]$ & $1.95^{+3.30}_{-0.78}\times 10^{-3}$ \\
$\epsilon_{e,2}$ & log-uniform & $[10^{-3}, 0.5]$ & $5.37^{+1.39}_{-1.01}\times 10^{-2}$ \\
$\epsilon_{B,1}$ & log-uniform & $[5\times10^{-3}, 0.5]$ & $4.57^{+4.55}_{-3.40}\times 10^{-2}$ \\
$\epsilon_{B,2}$ & log-uniform & $[10^{-4}, 0.5]$ & $1.15^{+1.67}_{-0.50}\times 10^{-2}$ \\
\enddata
\tablecomments{The upper part of the table presents the results of the two-component jet model, and the lower part presents those of the collisionless two-shell model.}
\end{deluxetable*}

\begin{deluxetable*}{ccccc}\label{tab_phot}
\digitalasset
\tablewidth{0pt}
\tablecaption{Optical photometry results of EP250302a \label{tab:description}}
\tablehead{
\colhead{Time (days)} & \colhead{Band} & \colhead{AB Magnitude} & \colhead{Telescope} & \colhead{Reference}
}
\startdata
1.03 & VT-B & $22.68 \pm 0.14$ & SVOM/VT & \citet{2025GCN.39578....1W} \\ 
0.0458 & UVOT-U & $18.61 \pm 0.05$ & SWIFT/UVOT & this work \\ 
0.0940 & UVOT-U & $19.82 \pm 0.12$ & SWIFT/UVOT & this work \\ 
1.59 & UVOT-U & $ \textgreater 22.40$ & SWIFT/UVOT & this work \\ 
3.56 & UVOT-U & $\textgreater 22.66$ & SWIFT/UVOT & this work \\ 
8.53 & UVOT-U & $\textgreater 22.46$ & SWIFT/UVOT & this work \\ 
0.0378 & g & $17.85 \pm 0.02$ & XL/TNT & this work \\ 
0.0544 & g & $18.12 \pm 0.08$ & NanShan/PAT17 & this work \\ 
0.0655 & g & $18.91 \pm 0.11$ & NanShan/PAT17 & this work \\ 
0.0765 & g & $19.33 \pm 0.12$ & NanShan/PAT17 & this work \\ 
0.0788 & g & $19.08 \pm 0.03$ & NUTTelA-TAO & this work \\ 
0.0875 & g & $19.47 \pm 0.13$ & NanShan/PAT17 & this work \\ 
0.0948 & g & $19.49 \pm 0.05$ & NUTTelA-TAO & this work \\ 
0.113 & g & $19.67 \pm 0.06$ & NUTTelA-TAO & this work \\ 
0.127 & g & $20.13 \pm 0.30$ & NanShan/PAT17 & this work \\ 
0.135 & g & $20.09 \pm 0.09$ & NUTTelA-TAO & this work \\ 
0.189 & g & $20.45 \pm 0.17$ & XL/TNT & this work \\ 
0.207 & g & $20.34 \pm 0.21$ & NanShan/PAT17 & this work \\ 
0.242 & g & $20.66 \pm 0.10$ & GROWTH-India Telescope & \citet{2025GCN.39575....1E} \\ 
0.00886 & r & $18.60 \pm 0.09$ & NanShan/HMT & this work \\ 
0.00992 & r & $18.85 \pm 0.12$ & NanShan/HMT & this work \\ 
0.0110 & r & $18.56 \pm 0.08$ & NanShan/HMT & this work \\ 
...\\
\enddata
\tablecomments{The magnitudes have not been corrected for Galactic extinction.}
\tablecomments{The full tables are available in the HTML version.}
\end{deluxetable*}

\bibliography{refs}{}
\bibliographystyle{aasjournalv7}



\end{document}